\documentclass[letter,11pt]{article}
\pdfoutput=1
\usepackage{jheppub}
\usepackage[dvipsnames,table,xcdraw]{xcolor}
\usepackage[T1]{fontenc}
\usepackage[colorlinks=true, linkcolor=blue, urlcolor=blue, citecolor=blue, anchorcolor=blue]{hyperref}
\usepackage{subcaption}

\newcommand{\cD}{\ensuremath{\mathcal D} }

\newcommand{\cN}{\ensuremath{\mathcal N} }
\newcommand{\cO}{\ensuremath{\mathcal O} }

\newcommand{\cP}{\ensuremath{\mathcal P} }
\newcommand{\That}{\ensuremath{\widehat{T}} }
\newcommand{\Tc}{\ensuremath{T_{\text{c}}} }
\newcommand{\Thatc}{\ensuremath{\That_{\text{c}}} }
\newcommand{\al}{\ensuremath{\alpha} }
\newcommand{\be}{\ensuremath{\beta} }
\newcommand{\ga}{\ensuremath{\gamma} }

\newcommand{\de}{\ensuremath{\delta} }
\newcommand{\De}{\ensuremath{\Delta} }
\newcommand{\eps}{\ensuremath{\epsilon} }
\newcommand{\la}{\ensuremath{\lambda} }
\newcommand{\lalat}{\ensuremath{\la_{\text{lat}}} }
\newcommand{\muhat}{\ensuremath{\widehat{\mu}} }
\newcommand{\mulat}{\ensuremath{\mu_{\text{lat}}} }
\newcommand{\si}{\ensuremath{\sigma} }
\newcommand{\om}{\ensuremath{\omega} }
\newcommand{\nn}{\nonumber }

\newcommand{\SO}[1]{\ensuremath{\text{SO(}#1\text{)}} }

\newcommand{\vev}[1]{\ensuremath{\left\langle #1 \right\rangle} }
\newcommand{\eq}[1]{Eq.~(\ref{#1})}
\newcommand{\fig}[1]{Fig.~\ref{#1}}
\newcommand{\tab}[1]{Table~\ref{#1}}
\newcommand{\secref}[1]{Sec.~\ref{#1}}
\newcommand{\refcite}[1]{Ref.~\cite{#1}}

\title{Non-perturbative phase structure of the bosonic BMN matrix model}

\author[a]{Navdeep~Singh~Dhindsa,}
\author[b]{Raghav~G.~Jha,}
\author[a]{Anosh~Joseph,}
\author[a]{Abhishek~Samlodia,}
\author[c]{David~Schaich}

\affiliation[a]{Department of Physical Sciences, Indian Institute of Science Education and Research - Mohali, Knowledge City, Sector 81, SAS Nagar, Punjab 140306, India}
\affiliation[b]{Perimeter Institute for Theoretical Physics, Waterloo, Ontario N2L 2Y5, Canada}
\affiliation[c]{Department of Mathematical Sciences, University of Liverpool, Liverpool L69 7ZL, United Kingdom}

\emailAdd{navdeep.s.dhindsa@gmail.com}
\emailAdd{raghav.govind.jha@gmail.com}
\emailAdd{anoshjoseph@iisermohali.ac.in}
\emailAdd{abhishek.s.samlodia@gmail.com}
\emailAdd{david.schaich@liverpool.ac.uk}

\abstract{We study the bosonic part of the BMN matrix model for wide ranges of temperatures, values of the deformation parameter, and numbers of colors $16 \leq N \leq 48$.
  Using lattice computations, we analyze phase transitions in the model, observing a single first-order transition from a uniform to a gapped phase for all values of the deformation parameter.
  We study the functional form of the dependence of the critical temperature on the deformation parameter, to describe how our results smoothly interpolate between the limits of the bosonic BFSS model and the gauged Gaussian model.
}

\begin{document}
\maketitle
\flushbottom

\section{Introduction}
\label{sec:intro}

One of the major developments in various attempts to understand the features of quantum gravity was the observation that the lower-dimensional models of matrices can capture the dynamics of string/M-theory in an appropriate limit of the parameters.
One of the first examples was the relation shown by \refcite{Klebanov:1991qa} between the (non-supersymmetric) $(0+1)$-dimensional $c = 1$ matrix model and two-dimensional bosonic string theory.
This program of connecting quantum-mechanical models to string/M-theory was extended by Banks, Fischler, Susskind, and Shenker (BFSS) through their proposal that the dimensional reduction of ten-dimensional $\cN = 1$ super-Yang--Mills (SYM) with gauge group SU($N$) describes M-theory in the light-cone gauge, in the large-$N$ planar limit~\cite{Banks:1996vh}. 
A few years later, Berenstein, Maldacena, and Nastase~\cite{Berenstein:2002jq} extended this model by introducing a supersymmetry-preserving one-parameter deformation. The result, known as the BMN matrix model, describes a certain limit of Type~II string theory on a pp-wave background rather than the flat spacetime relevant for the BFSS model.

Though there has been excellent progress in understanding and verifying the gauge/gravity duality conjecture by studying $\cN = 4$ SYM in four dimensions using ideas of integrability, the lower-dimensional non-conformal analogs of the four-dimensional theory have not attracted as much attention.
Only a handful of analytical attempts using certain approximations have been made so far~\cite{Kabat:1999hp, Wiseman:2013cda}. Since it is difficult to verify the duality conjecture in the finite-temperature setting relevant for these cases, we need a method that can provide information about their strongly coupled regimes. This opens up the possibility of exploring the dual field theories using the ideas and tools of lattice field theory.
In this regard, there has been good progress in understanding both various aspects of $(0+1)$-dimensional matrix models as well as the thermodynamics of stacks of D$p$ branes with $p = 1$ and $2$, using the $(p+1)$-dimensional dual supersymmetric theories in Euclidean lattice spacetimes~\cite{Catterall:2008yz, Catterall:2010gf, Kadoh:2015mka, Filev:2015hia, Berkowitz:2016tyy, Berkowitz:2016jlq, Catterall:2017lub, Jha:2017zad, Asano:2018nol, Schaich:2018mmv, Catterall:2020nmn, Bergner:2021goh, Schaich:2022duk, Sherletov:2022rnl}.\footnote{We thank Jun Nishimura and Masanori Hanada for comments on non-lattice numerical studies in $0+1$ dimensions~\cite{Anagnostopoulos:2007fw, Hanada:2008gy, Hanada:2008ez, Hanada:2009ne, Hanada:2011fq, Hanada:2013rga, Hanada:2016zxj}.}

One of the striking features of the gauge/gravity duality is that at large $N$ and finite temperature, there are often phase transitions between different quantum black hole solutions, which are dual to confinement transitions in the field theory. 
In this regard, the D$0$ brane matrix model is an exception, with only a single deconfined phase at all temperatures in the planar limit. 
But this behavior is drastically altered if we consider either a one-parameter deformation of the BFSS model, i.e., the BMN model, or if we decouple the fermions and study the bosonic sector of the BFSS model. In both cases, there is a well-defined confinement transition. The dual black hole solutions of the BMN model in the deconfined phase and the details of the phase transition were studied in \refcite{Costa:2014wya}. It remains a challenge to understand the phase diagram for finite couplings and to verify the results obtained using gravity computations.
Refs.~\cite{Asano:2018nol, Bergner:2021goh, Schaich:2022duk} have numerically explored the phase structure of the full BMN model, with \refcite{Asano:2018nol} reporting two different phase transitions --- a confinement transition signalled by the Polyakov loop, and a `Myers transition' signalled by the trilinear `Myers term' --- which merge into one for the dimensionless BMN deformation parameter $\muhat \equiv \mu / \la^{1/3} \lesssim 3$ (where \la is the dimensionful 't~Hooft coupling).
When these transitions are distinct, \refcite{Asano:2018nol} observes the Myers transition to be between two deconfined phases, one where the system fluctuates around the trivial configuration and the other with fluctuations around expanded fuzzy spheres.
More recently, as our work was in progress, \refcite{Bergner:2021goh} revisited the phase structure of the BMN model, introducing constraints on the Myers term to suppress fuzzy sphere contributions and focus on the confinement transition.

In this work, we report a detailed study of the phase structure that results upon both including the BMN deformation and decoupling the fermions.
While removing the fermions completely eliminates the $Q = 16$ supersymmetries of the BMN model, and its holographic connection to quantum gravity, we take this step in order to accelerate numerical computations.
This makes it easier to study larger lattice sizes and $N$, and thereby obtain more precise and reliable results for the phase structure.
These robust results will provide a solid starting point for subsequent efforts to analyze the full supersymmetric theory.

This bosonic BMN model was investigated in \refcite{Kovacik:2020cod, Asano:2020yry} for a fixed $\muhat = 2$, finding a single first-order transition in the large-$N$ limit, at the dimensionless critical temperature $\Thatc \equiv T / \la^{1/3} = 0.915(5)$.
It was not clear from this work whether different \muhat values might exhibit a Myers transition distinct from the confinement transition that was previously reported for the full BMN model by \refcite{Asano:2018nol}.
Addressing this question is part of the motivation for our investigations.
The recent \refcite{Bergner:2021goh} appeared while our work was underway, reporting a single first-order phase transition for $0.375 \leq \muhat \leq 3$ (the range where the two transitions had merged for the full BMN model~\cite{Asano:2018nol}).\footnote{Our \muhat is equivalent to $3\mu$ in the conventions of \refcite{Bergner:2021goh}.}
We will push further into the large-$\muhat$ regime, which will allow us to conclude that the bosonic BMN model features a single first-order transition for \emph{all} values of the deformation parameter.

Our goal is to explore the functional form of the dependence of the bosonic BMN critical temperature \Thatc on the deformation parameter $\muhat$.
To this end, we analyze twelve different values of \muhat spanning two orders of magnitude between the previously studied $\muhat \to 0$ and $\muhat \to \infty$ limits.
As $\muhat \to 0$, we recover the bosonic version of the BFSS model.
Although early numerical and analytic bosonic BFSS investigations reported two near-by phase transitions~\cite{Kawahara:2007fn, Mandal:2009vz}, more recent lattice calculations find only a single confinement transition with $\Thatc|_{\muhat = 0} = 0.8846(1)$~\cite{Bergner:2019rca, Bergner:2021goh}.
In the $\muhat \to \infty$ limit the system reduces to a solvable gauged Gaussian model, and for large \muhat the critical temperature is found to scale as $\Thatc = (6 \ln (3 + 2 \sqrt{3}))^{-1} \muhat$~\cite{Aharony:2003sx, Furuuchi:2003sy}.
These two limits serve as consistency checks for our lattice computations.

The paper is organized as follows. In \secref{sec:lat_for}, we discuss the lattice formulation and define the relevant observables we study.
In \secref{sec:lattice_results}, we present our results for a wide range of $0.5 \leq \muhat \lesssim 45$ with $N = 16$, $32$, and $48$.
The data leading to these results are available through \refcite{data}.
We additionally discuss a \emph{separatrix} method that we employ as a novel means to precisely estimate the critical temperature.
We then study the \muhat dependence of these critical temperatures, fitting them to different functional forms for small and large $\muhat$.
In \secref{sec:summary} we summarize our results and discuss next steps.

\section{Bosonic BMN model on a lattice}
\label{sec:lat_for}

The BMN model is a one-parameter deformation of the BFSS model---the dimensional reduction of $(9+1)$-dimensional $\cN = 1$ SYM with gauge group SU($N$) down to $0+1$ dimensions. In Euclidean time the action of the BFSS model is
\begin{equation}
  \label{eqn:s_bfss}
  \begin{split}
    S_{\text{BFSS}} & = \frac{N}{4 \la} \int_0^\be d\tau \ \mbox{Tr} \Big\{ -\left(D_{\tau} X_i\right)^2 - \frac{1}{2} \sum_{i < j} \left[X_i, X_j\right]^2 \\
    & \hspace{5 cm} + \Psi_{\al}^T \ga_{\al\si}^\tau D_\tau \Psi_{\si} + \Psi_{\al}^T \ga_{\al\si}^i \left[X_i, \Psi_{\si}\right] \Big\},
  \end{split}
\end{equation}
where $D_{\tau} \cdot = \partial_{\tau} \cdot + [A_{\tau}, \cdot]$ is the covariant derivative, $X_i$ are the nine scalars from the reduction of the ten-dimensional gauge field, and $\Psi_{\al}$ is a sixteen-component spinor. 
The indices $i, j = 1, \cdots, 9$ while $\al, \si = 1, \cdots, 16$. The degrees of freedom transform in the adjoint representation of the SU($N$) gauge group. The anti-Hermitan gauge group generators are normalized as $\mbox{Tr} (T^A T^B) = - \de_{AB}$. The trace `Tr' is taken over the gauge indices. In this $(0+1)$-dimensional model, the 't~Hooft coupling $\la \equiv g_{\text{YM}}^2 N$ is dimensionful, $[\la] = 3$. The model is compactified on a circle with circumference $\be = T^{-1}$, which corresponds to the inverse temperature because we impose thermal boundary conditions --- periodic for the bosons and anti-periodic for the fermions.

The action of the BMN model is obtained by adding the following mass and scalar-trilinear terms to \eq{eqn:s_bfss}:
\begin{equation}
S_{\mu} = -\frac{N}{4 \la} \int_0^\be d\tau \ \mbox{Tr} \left[ \left(\frac{\mu}{3} X_I \right)^2 + \left(\frac{\mu}{6} X_A \right)^2 + \frac{\mu}{4} \Psi_{\al}^T \ga_{\al \si}^{123} \Psi_{\si} - \frac{\sqrt{2} \mu}{3} \eps_{IJK} X_I X_J X_K \right].
\label{eq:deform}
\end{equation}
Here $\mu$ is the deformation parameter, with dimension $[\mu] = 1$. We divide the indices $i, j$ into two sets: $I, J, K = 1, 2, 3$ and $A = 4, \cdots, 9$.
The scalar mass terms break the $\SO{9}$ global symmetry of the BFSS model down to $\SO{3} \times \SO{6}$.
As $\mu \to \infty$ the model reduces to a free supersymmetric Gaussian model, and it can be studied perturbatively for large $\mu$~\cite{Dasgupta:2002hx, Dasgupta:2002ru}.
Since we are interested only in the bosonic sector, we can remove the fermions to obtain the action of the bosonic BMN (BBMN) model:
\begin{equation}
  \label{eq:action}
  \begin{split}
    S_{\text{BBMN}} & = \frac{N}{4 \la} \int_0^{\be} d\tau \ \mbox{Tr} \Bigg[ - \left( D_{\tau} X_i \right)^2 - \frac{1}{2} \sum_{i < j} \left[X_i, X_j\right]^2 \\
    & \hspace{5 cm}  - \left( \frac{\mu}{3} X_I \right)^2 - \left( \frac{\mu}{6} X_A \right)^2 + \frac{\sqrt{2} \mu}{3} \eps_{IJK} X_I X_J X_K \Bigg].
  \end{split}
\end{equation}

We discretize this model on a lattice with $N_{\tau}$ sites. The inverse temperature becomes $\be = a N_{\tau}$, where `$a$' is the lattice spacing with dimension $[a] = -1$. The integration becomes a summation over the lattice sites: $\int_0^{\be} d\tau \longrightarrow a \sum_{n = 0}^{N_{\tau} - 1}$. The dimensionful gauge field $A_{\tau}$ is mapped to a dimensionless gauge link $U(n)$ connecting the sites $n$ and $n+1$. We also work with dimensionless scalars $X_i(n) = aX_i(\tau)$ once we are on the lattice. To discretize the covariant derivative $D_{\tau} X_i(\tau)$ we use the gauge link to define the finite-difference operator $\cD_+ X_i(n) \equiv U(n) X_i(n + 1) U^{\dag}(n) - X_i(n)$.
Finally, we introduce the dimensionless lattice parameters $\mulat \equiv a \mu$ and $\lalat \equiv a^3 \la$, to end up with a lattice action for the bosonic BMN model that has the same form as \eq{eq:action} while employing only dimensionless lattice quantities:
\begin{equation}
  \label{eq:lat_act}
  \begin{split}
    S_{\text{lat}} & = \frac{N}{4 \lalat} \sum_{n = 0}^{N_{\tau} - 1} \mbox{Tr} \Bigg[ - \left(\cD_+ X_i\right)^2 - \frac{1}{2} \sum_{i < j} \left[X_i, X_j\right]^2 \\
    & \hspace{4.5 cm}  - \left( \frac{\mulat}{3} X_I \right)^2 - \left( \frac{\mulat}{6} X_A \right)^2 + \frac{\sqrt{2} \mulat}{3} \eps_{IJK} X_I X_J X_K \Bigg].
  \end{split}
\end{equation}
The following dimensionless combinations of parameters are particularly useful, because they can be considered consistently in both the lattice and continuum theories:
\begin{align}
  \That & \equiv \frac{T}{\la^{1/3}} = \frac{1}{N_{\tau}\lalat^{1/3}} &
  \muhat & \equiv \frac{\mu}{\la^{1/3}} = \frac{\mulat}{\lalat^{1/3}} &
  \frac{\That}{\muhat} & = \frac{T}{\mu} = \frac{1}{N_{\tau} \mulat}.
\end{align}

Using this simple lattice action, we generate ensembles of matrix configurations using the hybrid Monte Carlo algorithm implemented by the publicly available parallel software presented in \refcite{Schaich:2014pda}.\footnote{\href{https://github.com/daschaich/susy}{github.com/daschaich/susy}}
As our main goal is to analyze phase transitions, we concentrate our lattice calculations around the transition regions.
We focus our analyses on the following four observables, again employing dimensionless quantities that connect smoothly between the continuum and lattice theories:
\begin{itemize}
\item The internal energy. As derived in Appendix~\ref{sec:int_energy}, on the lattice this is
\begin{equation}
  \label{eqn:E_b}
  \begin{split}
    \frac{\widehat{E}}{N^2} & \equiv \frac{E}{\la^{1/3} N^2} = \frac{1}{4 N \lalat^{4/3} N_\tau} \Bigg \langle \sum_{n = 0}^{N_\tau - 1} \mbox{Tr} \Bigg( - \frac{3}{2} \sum_{i < j} [X_i, X_j]^2 - \frac{2\mulat^2}{9} X_I^2 - \frac{\mulat^2}{18} X_A^2 \\
    & \hspace{7.5 cm} + \frac{5 \sqrt{2} \mulat}{6} \epsilon_{IJK} X^I X^J X^K \Bigg) \Bigg \rangle.
  \end{split}
\end{equation}

\item The scalar-trilinear term, also known as the Myers term. In the continuum, we define this as the dimensionless quantity
\begin{equation}
  \widehat{M} \equiv \frac{M}{\la} = \frac{\sqrt{2}}{12 N} \frac{1}{\la \be} \vev{\int d\tau \ \eps_{IJK} \ \mbox{Tr} \left(X_I X_J X_K \right)}.
\end{equation}
On the lattice, it takes the form
\begin{equation}
  \label{eq:Myers}
  \widehat{M} = \frac{\sqrt{2}}{12 N \lalat N_{\tau}} \vev{\sum_{n = 0}^{N_{\tau} - 1} \eps_{IJK} \ \mbox{Tr} \left(X_I X_J X_K \right)}.
\end{equation}

\item The Polyakov loop magnitude. The Polyakov loop is the holonomy around the time direction and is the order parameter for the confinement transition in the large-$N$ limit. We have
\begin{equation}
  |P| = \vev{\left\vert \frac{1}{N} \mbox{Tr} \ \cP \exp\left[- \int_0^{\be} d\tau \ A_\tau \right] \right\vert},
\end{equation}
where $\cP \exp$ is the path-ordered exponential.
Translating this to the lattice,
\begin{equation}
  \label{eq:Poly}
  |P| = \vev{\frac{1}{N} \left\vert \mbox{Tr} \left(\prod_{n = 0}^{N_{\tau} - 1} U(n)\right) \right\vert}.
\end{equation}

\item The `extent of space' --- terminology motivated by the holographic dual of the full BMN model --- which is given by the sum of the squared scalars.  In the continuum we consider the dimensionless quantity
\begin{align}
  \widehat{R}^2 \equiv \frac{R^2}{\la^{2/3}} = \frac{1}{2N \la^{2/3} \be} \vev{\int d\tau \ \mbox{Tr} \left(X_i^2 \right)}.
\end{align}
On the lattice this becomes
\begin{align}
  \label{eq:scalar_square}
  \widehat{R}^2 = \frac{1}{2N \lalat^{2/3} N_{\tau}} \vev{\sum_{n = 0}^{N_{\tau} - 1} \mbox{Tr} \left(X_i^2 \right)}.
\end{align}
\end{itemize}
In addition, to help identify and characterize transitions, we also consider two further observables related to those above:
\begin{itemize}
\item The susceptibility of the Polyakov loop magnitude,
\begin{equation}
  \label{eq:suscept}
  \chi \equiv N^2 \left(\vev{|P|^2} - \vev{|P|}^2 \right).
\end{equation}

\item The specific heat, which on the lattice takes the form
\begin{equation}
  C_{_V} \equiv \frac{\lalat^{2/3} N_{\tau}^2}{N^2} \vev{\left(\widehat{E} - \vev{\widehat{E}}\right)^2 - \widehat{E}'},
 \end{equation}
with $\widehat{E}$ from \eq{eqn:E_b} and
\begin{equation}
  \begin{split}
    \label{eq:spec}
    \widehat{E}' & = \frac{N}{4 \lalat^{5/3} N_{\tau}^2} \Bigg \langle \sum_{n = 0}^{N_{\tau} - 1} \mbox{Tr} \Bigg( -3 \sum_{i < j} [X_i, X_j]^2 - \frac{2 \mulat^2}{9} X_I^2 - \frac{\mulat^2}{18} X_A^2 \\
    & \hspace{6.5 cm} + \frac{5 \sqrt{2} \mulat}{4} \eps_{IJK} X^I X^J X^K \Bigg) \Bigg \rangle.
  \end{split}
\end{equation}
\end{itemize}
Considering both the Polyakov loop susceptibility and the specific heat will allow us to search separately for the confinement transition and the Myers transition reported by \refcite{Asano:2018nol} (for the full BMN model).
The latter is signalled by the energy (in addition to the Myers term), and hence by the specific heat.

In contrast to lattice studies of finite-temperature transitions in higher-dimensional field theories, the absence of any spatial volume in lattice discretizations of matrix models means that the thermodynamic limit corresponds to increasing the number of colors, $N \to \infty$.
Prior studies of supersymmetric matrix models~\cite{Berkowitz:2016tyy, Berkowitz:2016jlq} have found that $N \geq 16$ is needed to control finite-$N$ artifacts in this context~\cite{Schaich:2018mmv}.
We therefore consider $N = 16$, $32$ and $48$, finding that we need to increase $N$ as \muhat decreases, in order to keep finite-$N$ artifacts negligible compared to our statistical precision.

In addition to the $N \to \infty$ thermodynamic limit, lattice analyses also need to consider the $N_{\tau} \to \infty$ continuum limit.
Here we proceed by fixing $N_{\tau} = 24$, following preliminary $N = 16$ calculations for a range of lattice sizes, which indicated that $N_{\tau} = 24$ appears sufficient to keep finite-$N_{\tau}$ artifacts negligible compared to our statistical precision. 
This is also consistent with earlier lattice studies~\cite{Asano:2018nol, Bergner:2019rca, Kovacik:2020cod, Asano:2020yry}.

\section{Lattice results}
\label{sec:lattice_results}
\subsection{Determination of the critical temperature}

\begin{figure}[p]
  \centering
  \begin{subfigure}{0.475\textwidth}
    \includegraphics[width=\linewidth]{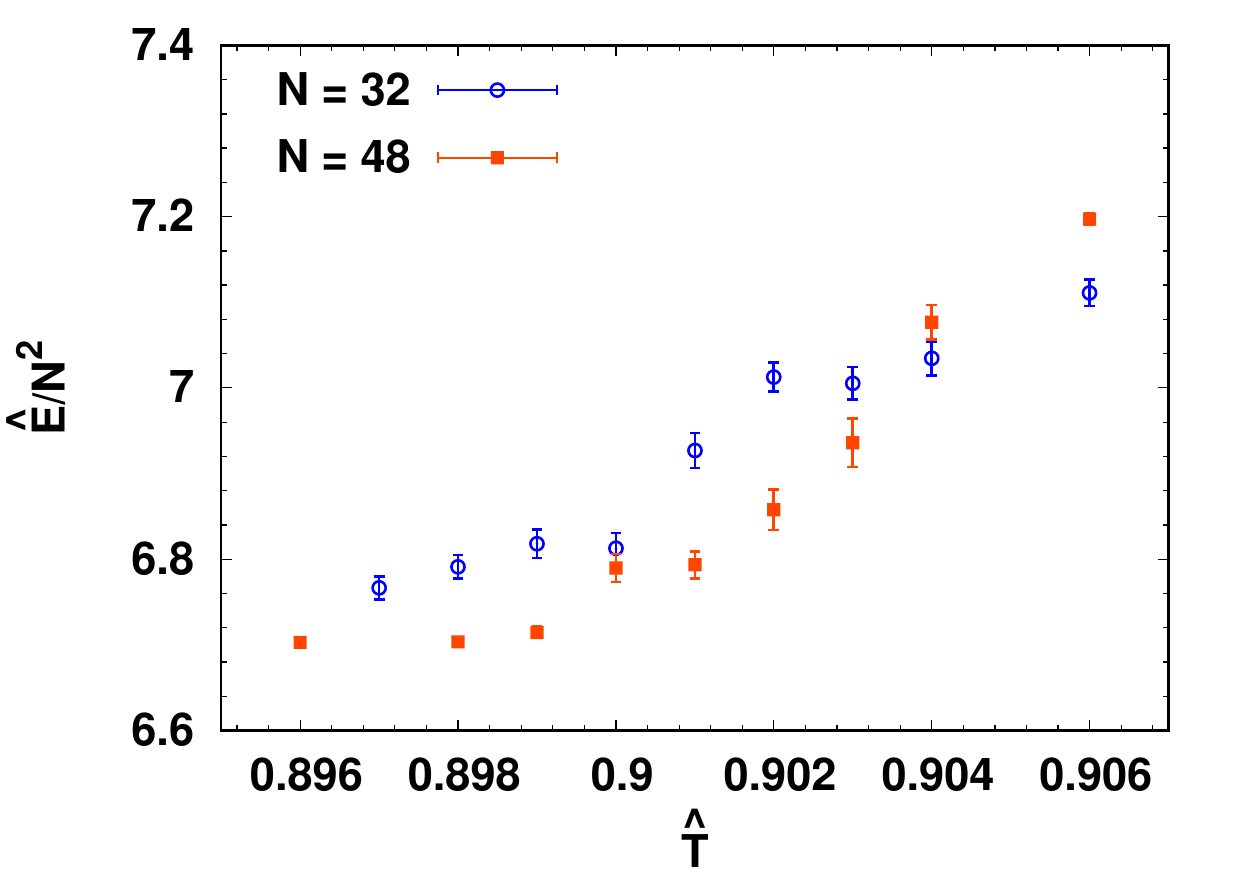}
    \caption{Internal energy, \eq{eqn:E_b}}
  \end{subfigure}
  \begin{subfigure}{0.475\textwidth}
    \includegraphics[width=1.0\linewidth]{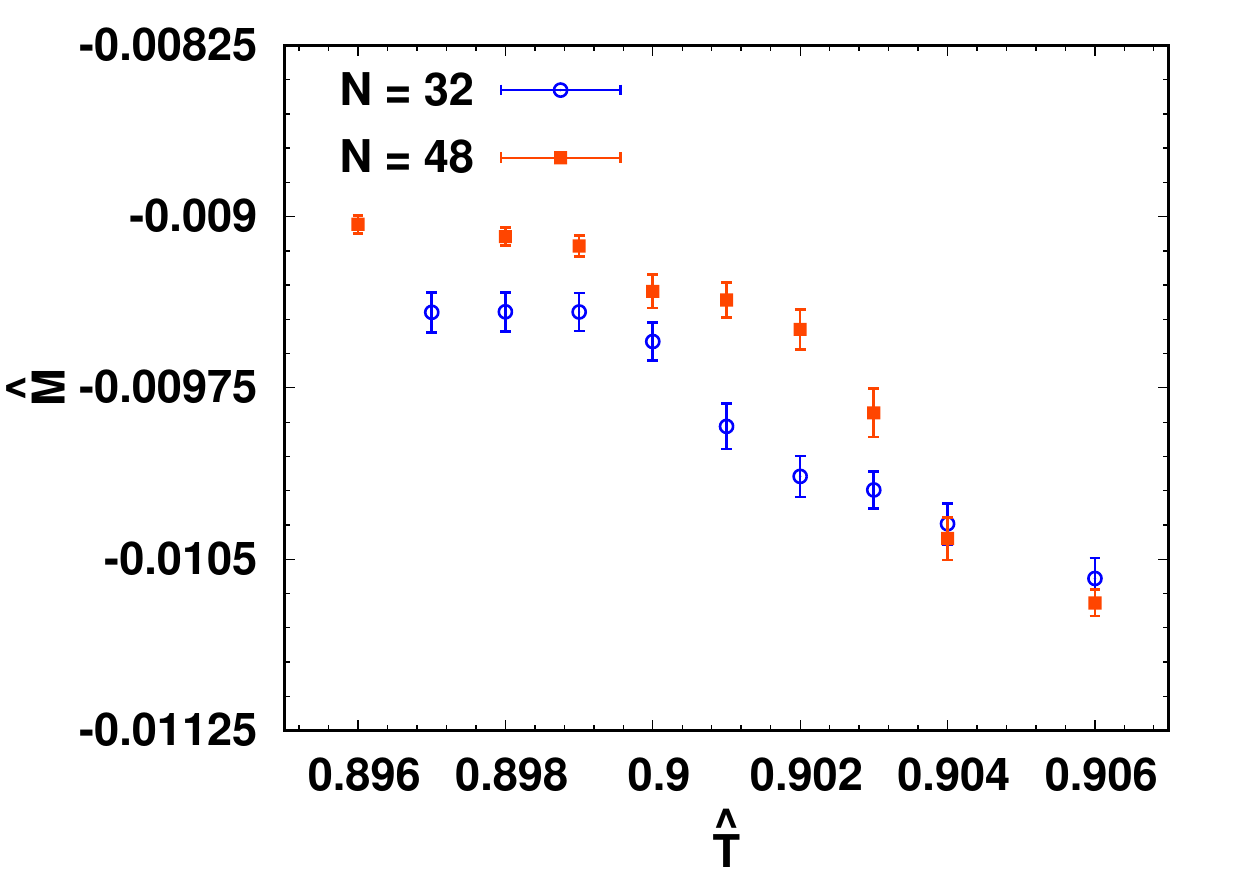}
    \caption{Myers term, \eq{eq:Myers}}
  \end{subfigure}
  \begin{subfigure}{0.475\textwidth}
    \includegraphics[width=\linewidth]{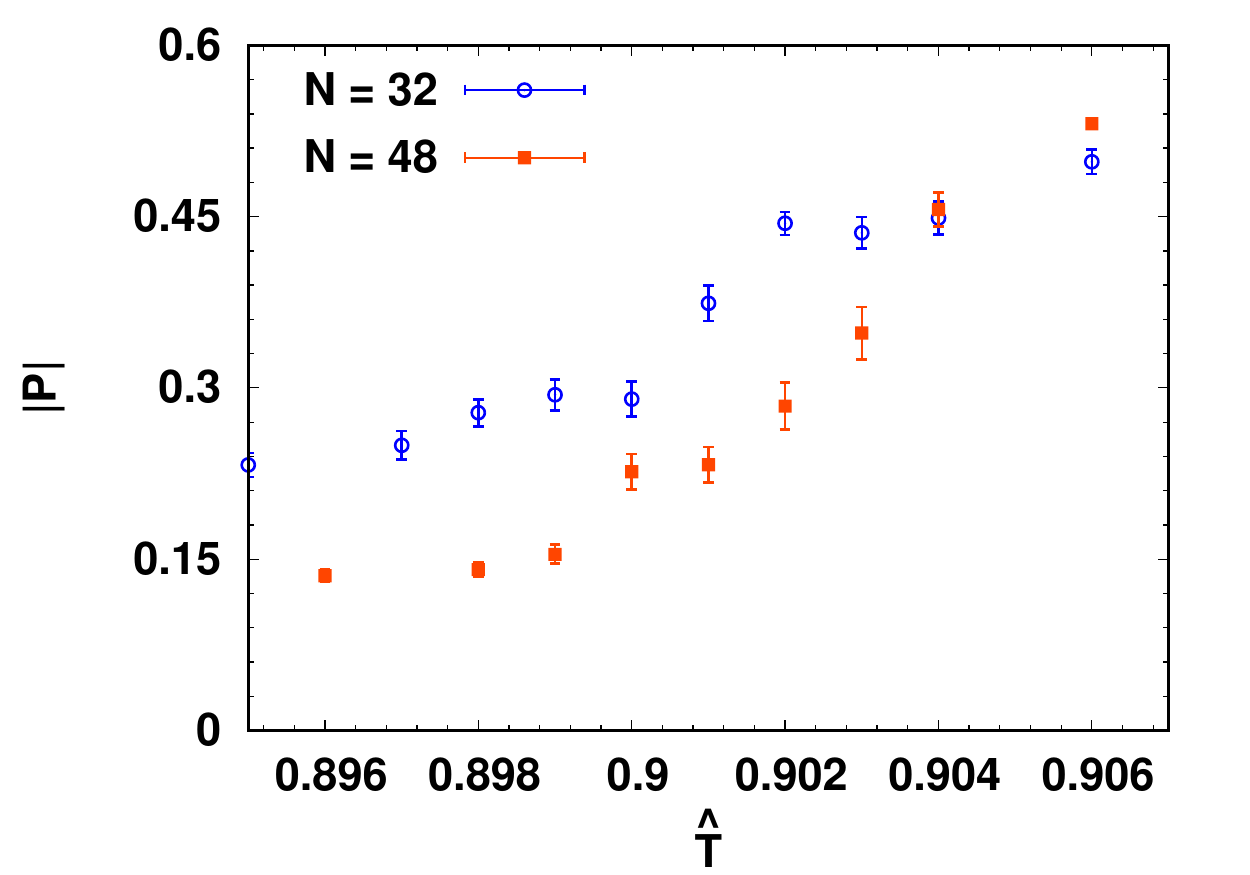}
    \caption{Polyakov loop, \eq{eq:Poly}}
  \end{subfigure}
  \begin{subfigure}{0.475\textwidth}
    \includegraphics[width=\linewidth]{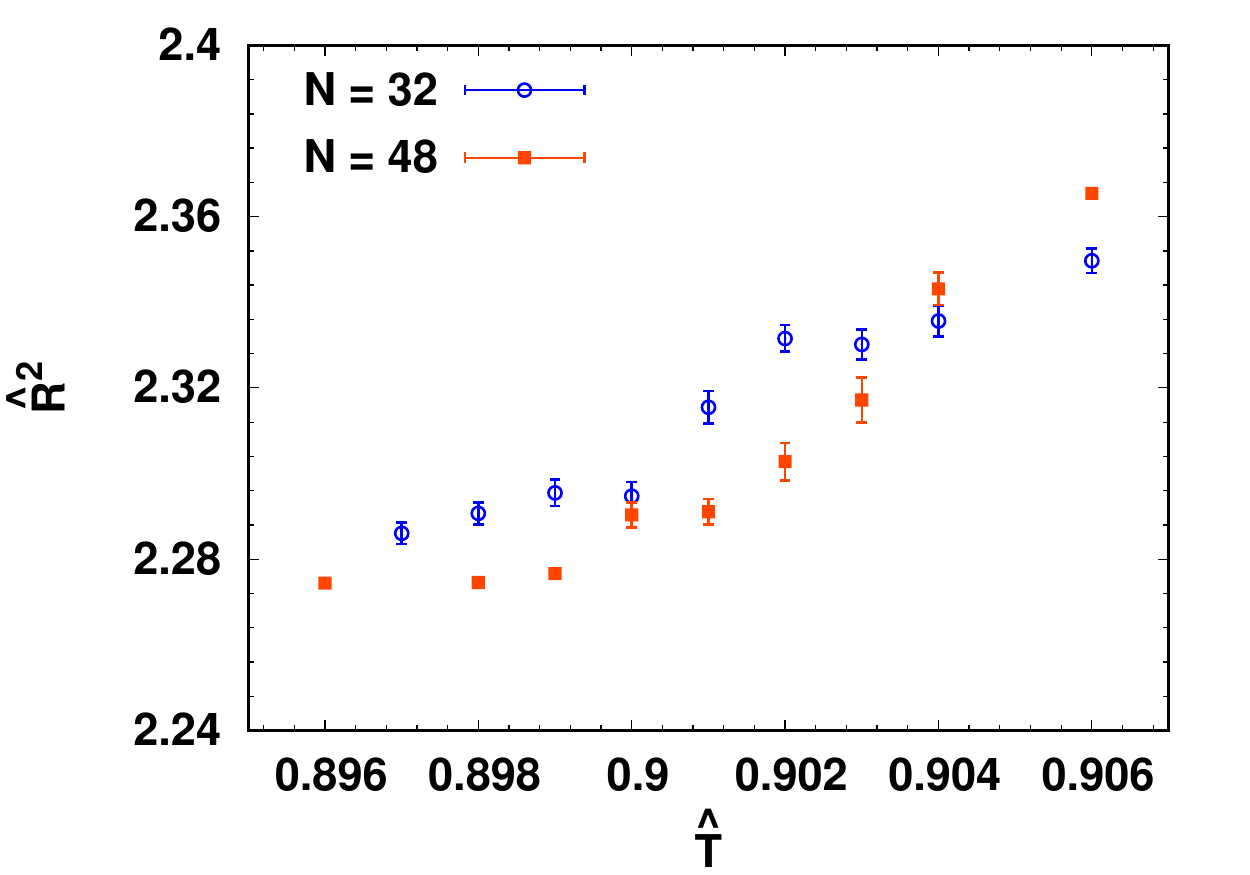}
    \caption{Extent of space, \eq{eq:scalar_square}}
  \end{subfigure}
  \begin{subfigure}{0.475\textwidth}
    \includegraphics[width=\linewidth]{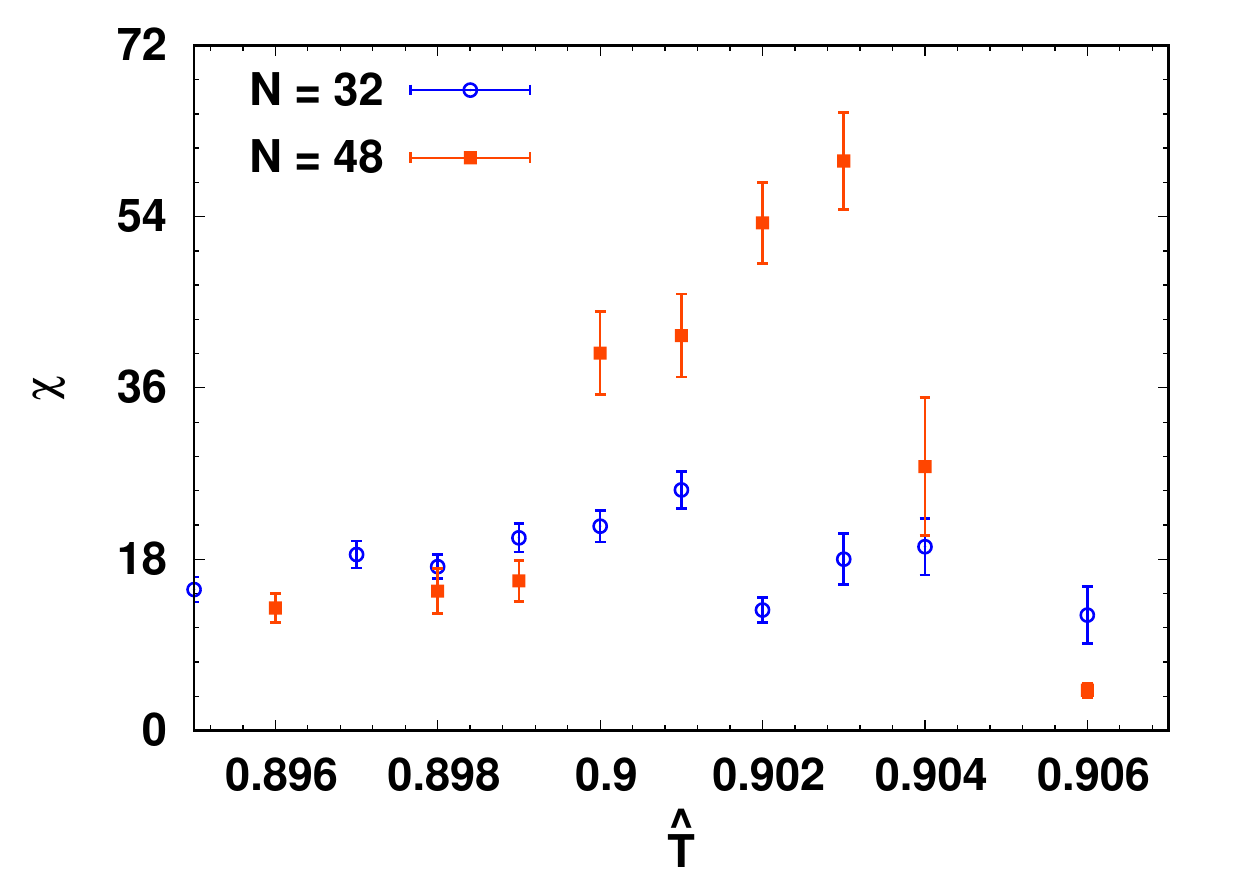}
    \caption{Polyakov loop susceptibility, \eq{eq:suscept}}
  \end{subfigure}
  \begin{subfigure}{0.475\textwidth}
    \includegraphics[width=\linewidth]{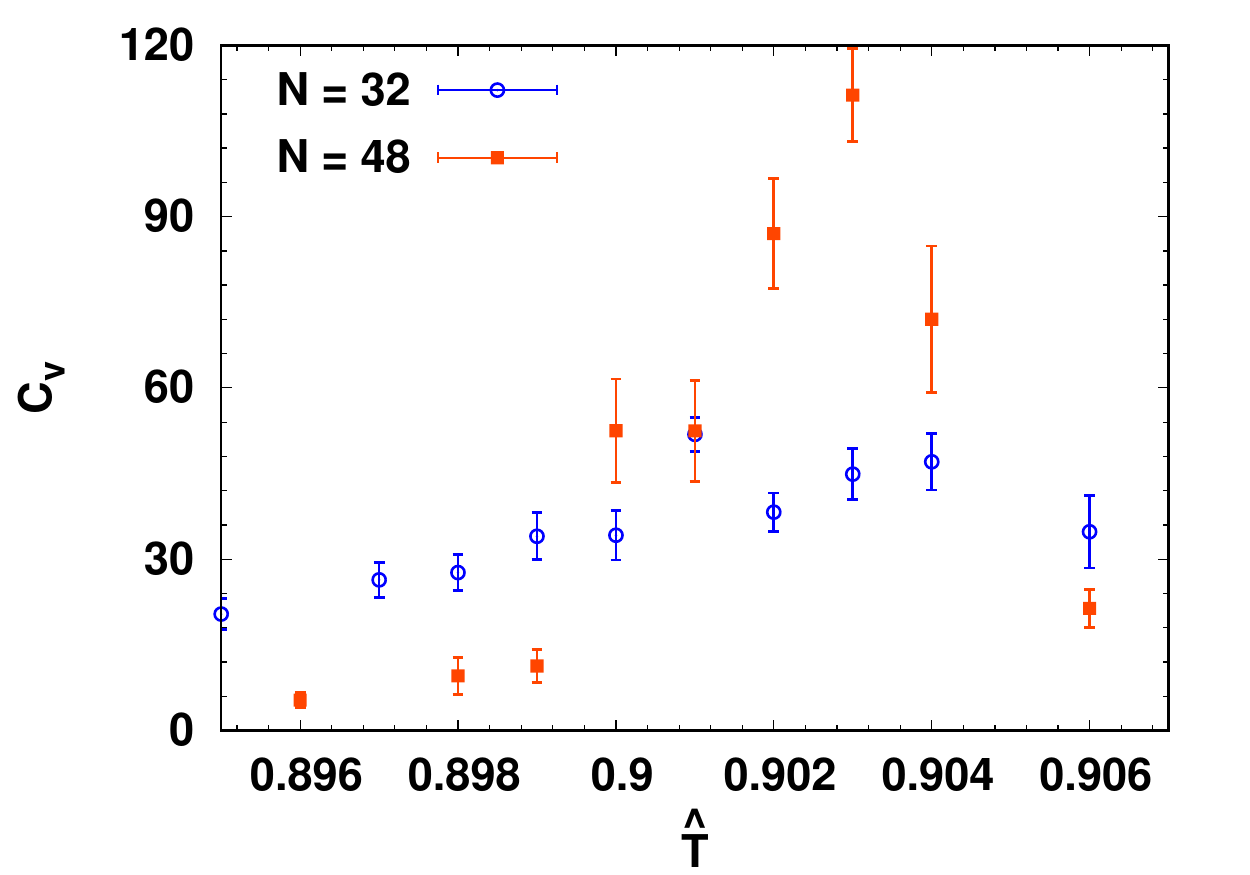}
    \caption{Specific heat, \eq{eq:spec}}
  \end{subfigure}
  \caption{\label{fig:N_dependence_mu1}The six observables discussed in the text, for $\muhat = 1$ and $N_{\tau} = 24$, in a small range of temperatures $0.896 \leq \That \leq 0.906$ around the transition.  As $N$ increases from $32$ to $48$ there is clear growth in the Polyakov loop susceptibility and the specific heat peaks in the final row.}
\end{figure}

\begin{figure}[p]
  \centering
  \begin{subfigure}{0.475\textwidth}
    \includegraphics[width=\linewidth]{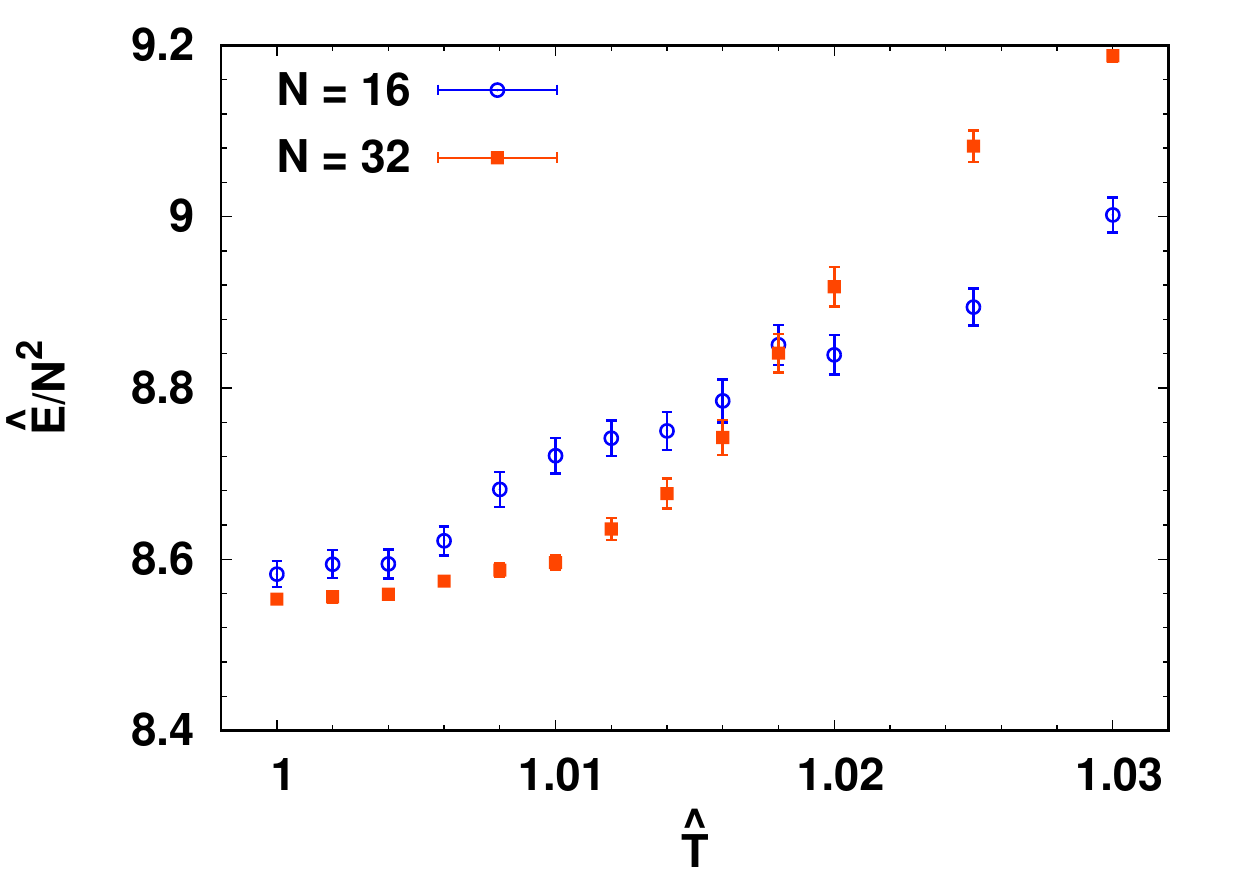}
    \caption{Internal energy, \eq{eqn:E_b}}
  \end{subfigure}
  \begin{subfigure}{0.475\textwidth}
    \includegraphics[width=1.0\linewidth]{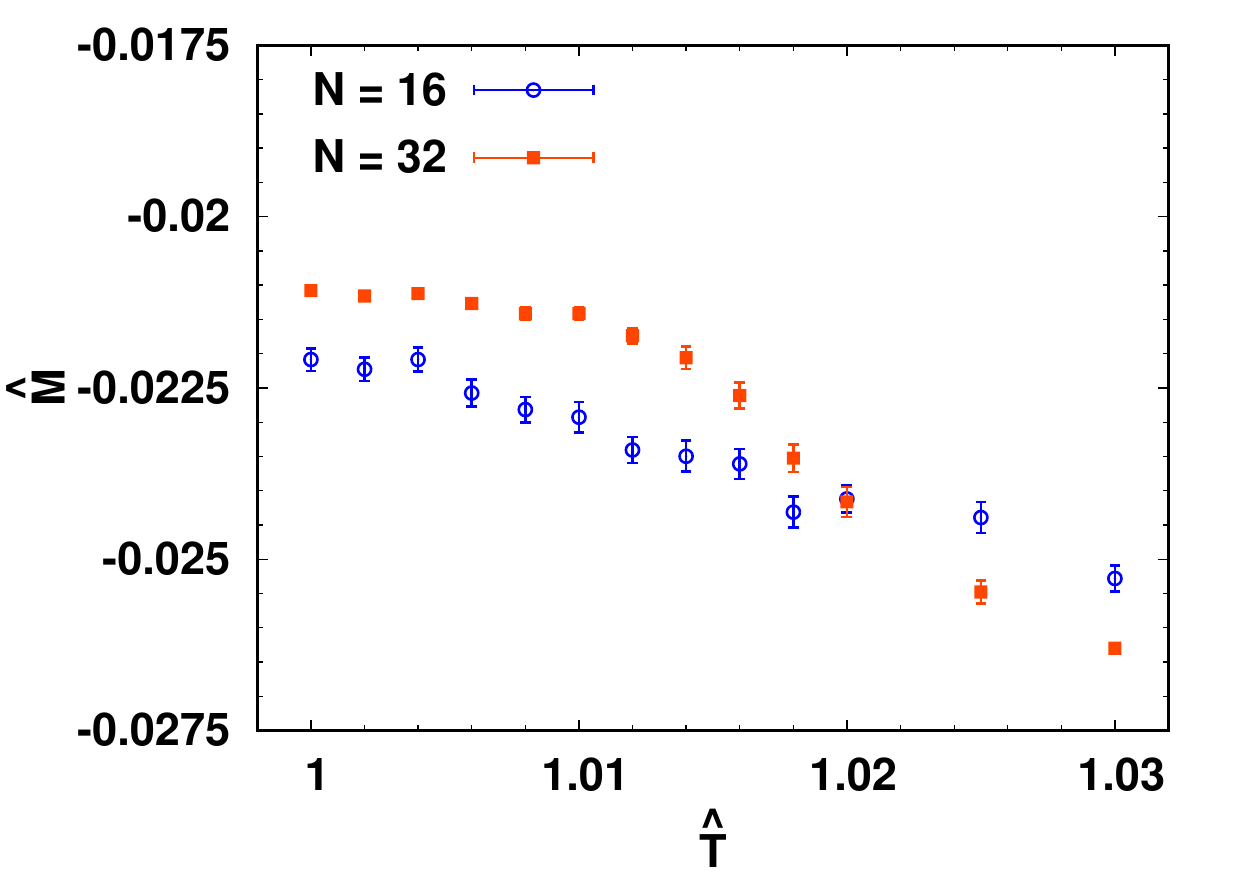}
    \caption{Myers term, \eq{eq:Myers}}
  \end{subfigure}
  \begin{subfigure}{0.475\textwidth}
    \includegraphics[width=\linewidth]{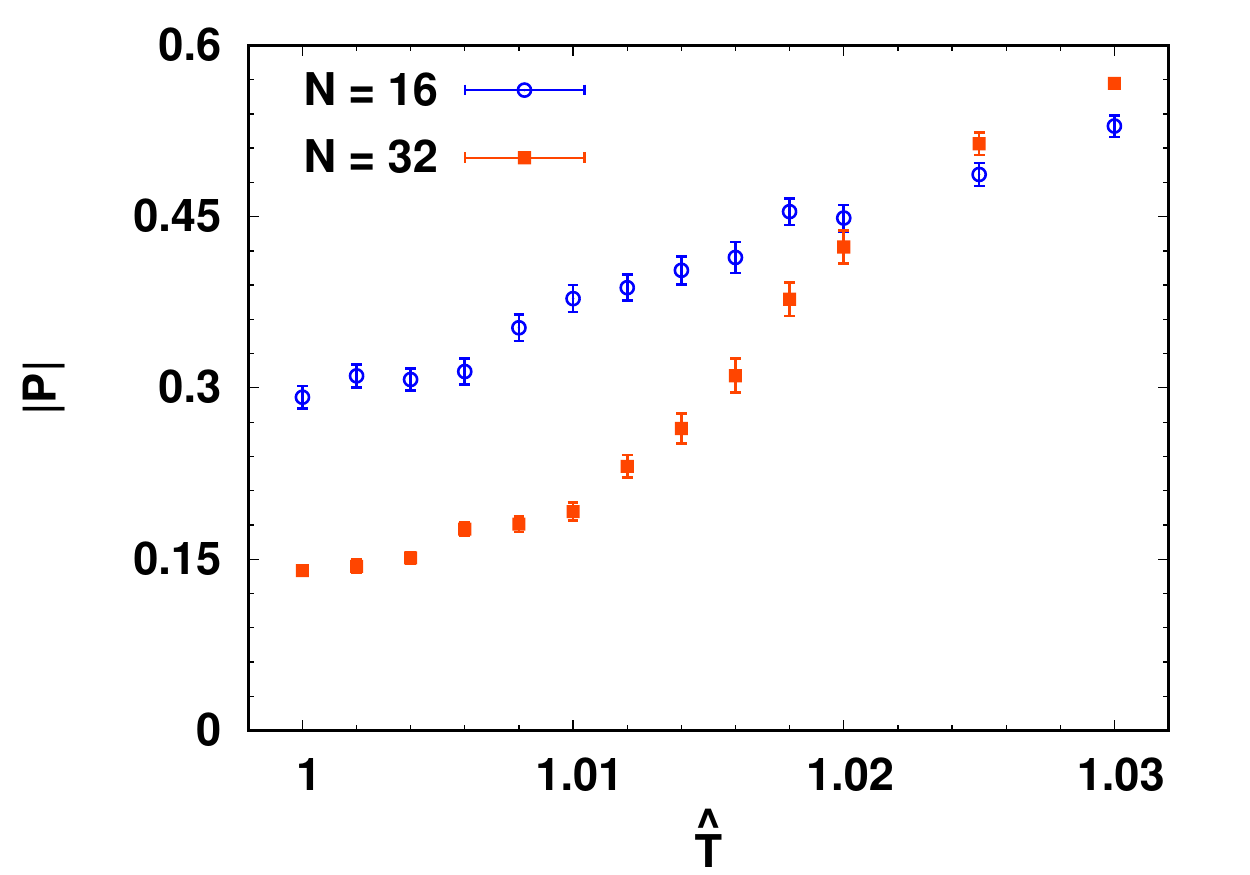}
    \caption{Polyakov loop, \eq{eq:Poly}}
  \end{subfigure}
  \begin{subfigure}{0.475\textwidth}
    \includegraphics[width=\linewidth]{Figures/Extent_mu1.pdf}
    \caption{Extent of space, \eq{eq:scalar_square}}
  \end{subfigure}
  \begin{subfigure}{0.475\textwidth}
    \includegraphics[width=\linewidth]{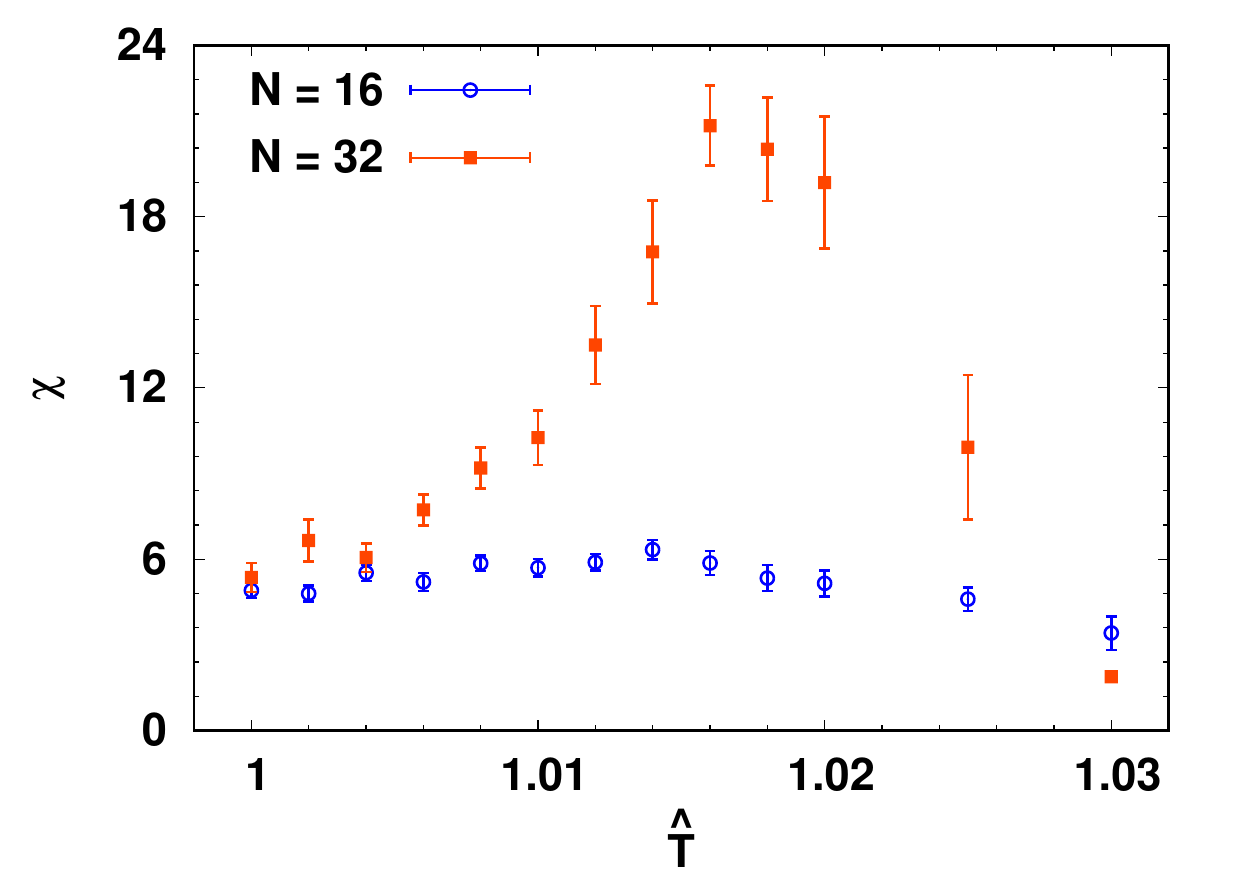}
    \caption{Polyakov loop susceptibility, \eq{eq:suscept}}
  \end{subfigure}
  \begin{subfigure}{0.475\textwidth}
    \includegraphics[width=\linewidth]{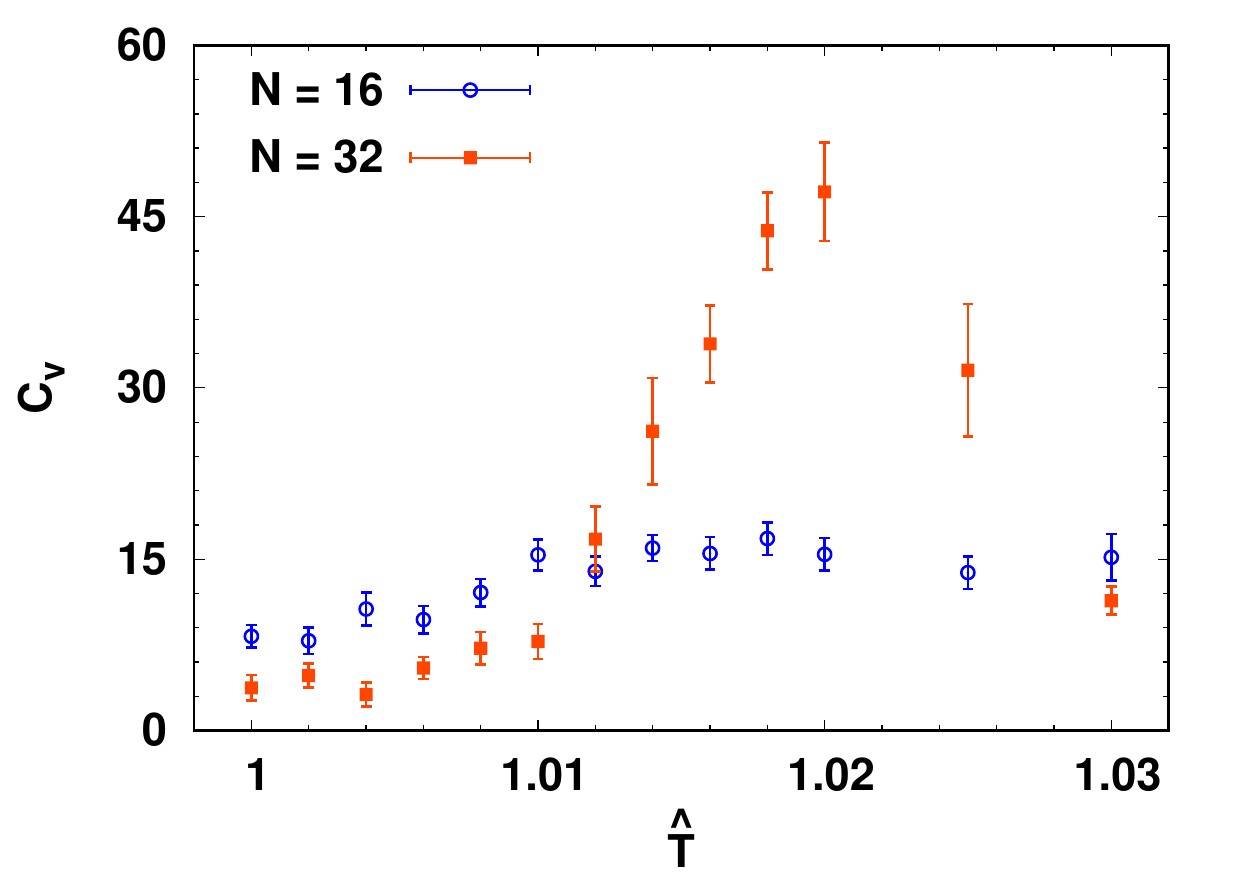}
    \caption{Specific heat, \eq{eq:spec}}
  \end{subfigure}
  \caption{\label{fig:N_dependence_mu6}The six observables discussed in the text, for $\muhat = 6$ and $N_{\tau} = 24$, in a small range of temperatures $1 \leq \That \leq 1.03$ around the transition.  As in \fig{fig:N_dependence_mu1}, increasing $N$ from $16$ to $32$ produces clear growth in the Polyakov loop susceptibility and the specific heat peaks in the final row.}
\end{figure}

As a first look at our lattice results, we collect some representative plots for the six observables summarized above.
In \fig{fig:N_dependence_mu1} we consider $\muhat = 1$ with $N_{\tau} = 24$, scanning the small range $0.896 \leq \That \leq 0.906$ around the transition and observing clear growth in the Polyakov loop susceptibility and specific heat peaks as the number of colors increases from $N = 32$ to $48$.
We see similar behavior in \fig{fig:N_dependence_mu6} for $\muhat = 6$ and $N_{\tau} = 24$ with $1 \leq \That \leq 1.03$, here comparing $N = 16$ and $32$.
Ref.~\cite{data} provides a comprehensive release of our data, including full accounting of statistics, auto-correlation times, and other observables computed in addition to the six highlighted in Figs.~\ref{fig:N_dependence_mu1} and \ref{fig:N_dependence_mu6}.

\begin{table}[t]
\centering
\renewcommand\arraystretch{1.2} 
\vspace{0.5cm}
\begin{tabular}{p{1.5cm}|p{1cm}|p{2cm}|p{2cm}}
\hline
\hline
\muhat & $N$ & \Thatc   & \De \\
\hline
 0.5   & 48  & 0.900(2) & 0.000(3) \\
 1.0   & 48  & 0.903(1) & 0.000(2) \\
 2.0   & 48  & 0.912(1) & 0.000(2) \\
 4.0   & 32  & 0.949(2) & 0.001(4) \\
 6.0   & 32  & 1.016(4) & 0.004(6) \\
 9.0   & 16  & 1.158(8) & 0.000(20) \\
10.0   & 16  & 1.213(8) & 0.010(26) \\
11.0   & 16  & 1.275(3) & 0.000(10) \\
13.0   & 16  & 1.398(8) & 0.012(18) \\
15.0   & 16  & 1.531(9) & 0.012(23) \\
21.54  & 16  & 2.04(3)  & 0.01(6) \\
44.66  & 16  & 4.00(3)  & -- \\
\hline
\end{tabular}
\caption{\label{tab:Tc}The critical temperature \Thatc for the $12$ \muhat values considered in this work.  As \muhat decreases, larger $N$ is needed.  We determine \Thatc either from the peak of the Polyakov loop susceptibility or from the separatrix method.  The last column shows the values of the \De parameter defined by \eq{eq:Delta}.}
\end{table}

Using $N_{\tau} = 24$, we performed similar scans in the temperature for all $12$ values of $0.5 \leq \muhat \leq 44.66$ listed in \tab{tab:Tc}.
In addition to identifying the critical temperature \Thatc from the peak in the Polyakov loop susceptibility, we also carry out analyses using the \emph{separatrix} method.
This is a novel way to determine the critical temperature away from the thermodynamic limit, which can work well even when susceptibility peaks are difficult to resolve.
While \refcite{Francis:2015lha} introduced the Polyakov loop separatrix method specifically for (non-supersymmetric) SU(3) Yang--Mills theory, it generalizes to $N > 3$.
The idea is to consider the unit disk in the plane of the real and imaginary parts of the Polyakov loop, and separate this into two regions such that Polyakov loop measurements for deconfined ensembles fall predominantly in one region while those from confined ensembles fall predominantly in the other.
A simple ratio $S(\That)$ then changes from $0$ deep in the deconfined phase to $1$ deep in the confined phase, with the transition identified as the (interpolated) point where this ratio crosses $0.5$.\footnote{\refcite{Kovacik:2020cod, Asano:2020yry} considers a similar ratio, but identifies the transition as the point where it becomes non-zero.}
Fig.~1 in \refcite{Francis:2015lha} illustrates the equilateral triangle used as the SU(3) separatrix.
For our $N \geq 16$ we approximate the corresponding $N$-gon by a circle of radius $r_S < 1$, so that we just have to consider the Polyakov loop magnitude $|P|$.

\begin{figure}[btp]
\centering
\begin{subfigure}{0.475\textwidth}
\includegraphics[width=\linewidth]{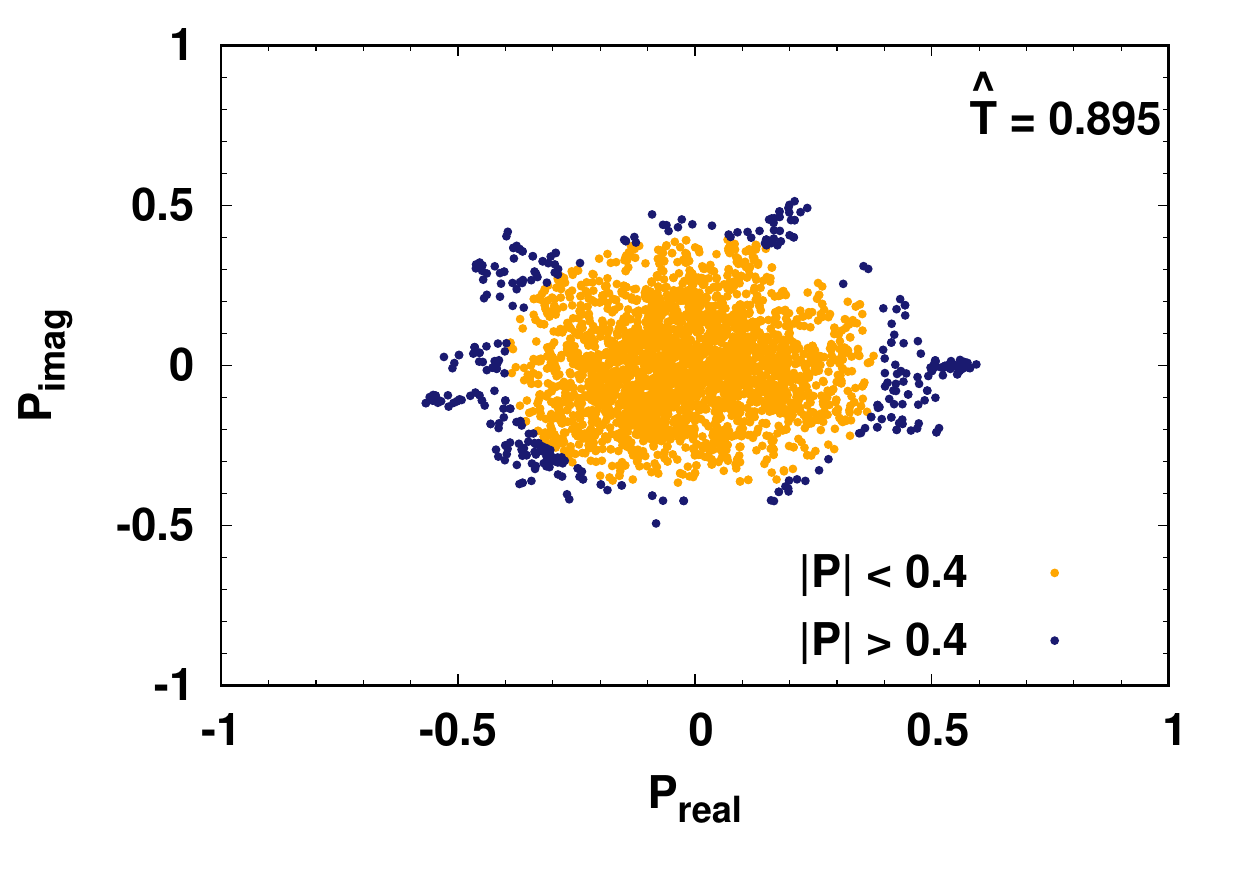}
\caption{$\That = 0.895$}
\end{subfigure}
\begin{subfigure}{0.475\textwidth}
\includegraphics[width=\linewidth]{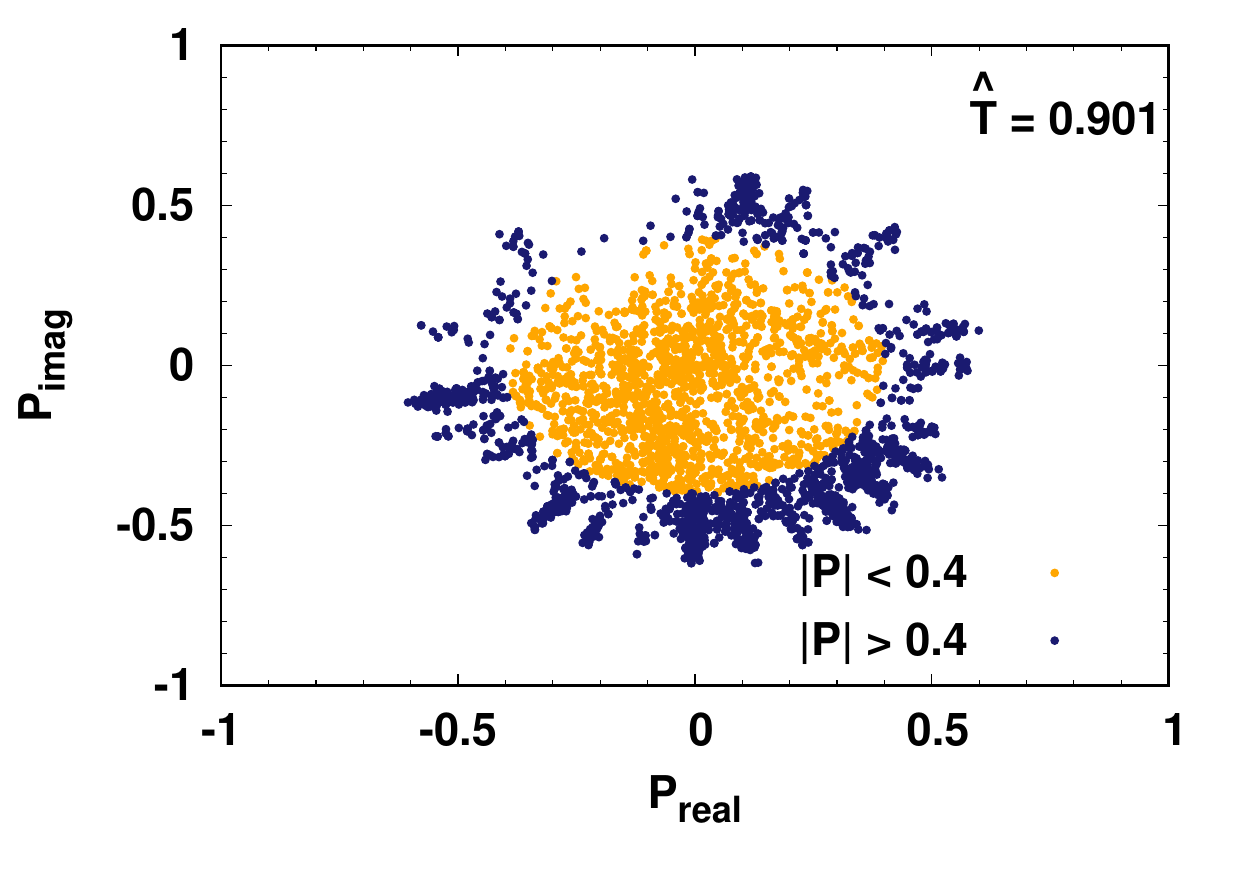}
\caption{$\That = 0.901$}
\end{subfigure}
\begin{subfigure}{0.475\textwidth}
\includegraphics[width=\linewidth]{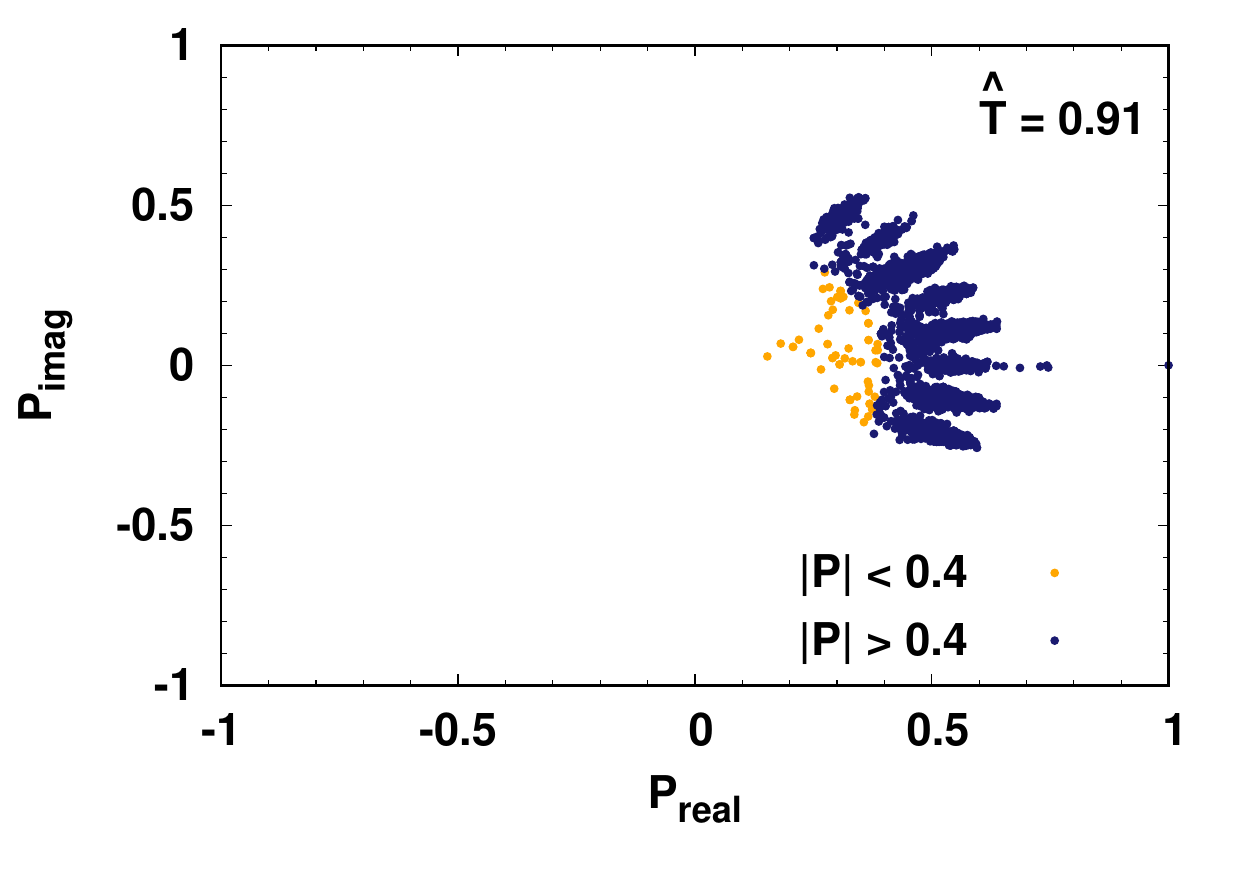}
\caption{$\That = 0.91$}
\end{subfigure}
\begin{subfigure}{0.475\textwidth}
\includegraphics[width=\linewidth]{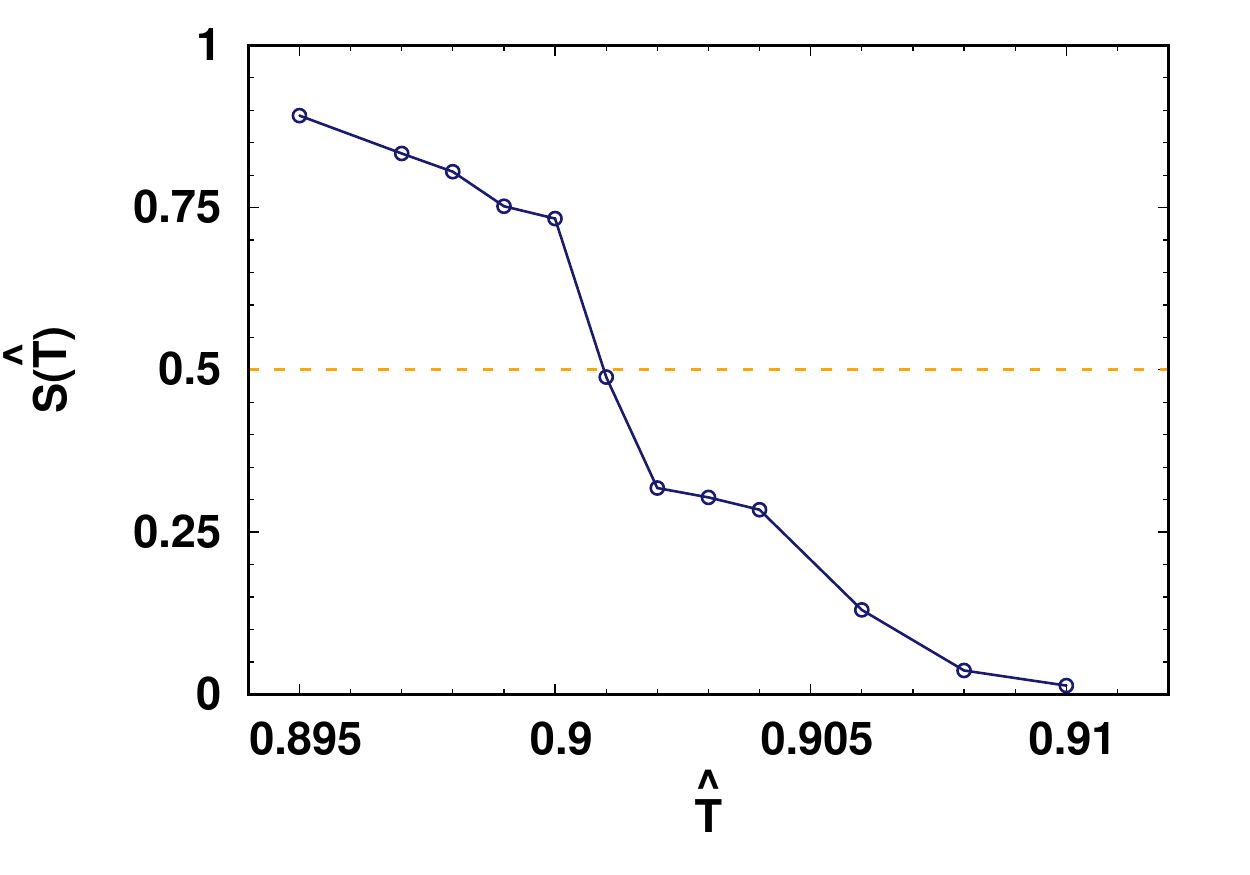}
\caption{$S(\That)$}
\end{subfigure}
\caption{\label{fig:scatter_separatrix_2}Polyakov loop scatter plots for $N = 32$ with $\muhat = 1$ for three values of $\That$.  The separatrix is a circle of radius $r_S = 0.4$ shown by the two colors.  Panel (d) shows the resulting $S(\That)$ that identifies the critical temperature $\Thatc \approx 0.901$.  (Note the $\Thatc = 0.903(1)$ in \tab{tab:Tc} comes from $N = 48$.)}
\end{figure}

\begin{figure}[tbp]
\centering
\begin{subfigure}{0.475\textwidth}
\includegraphics[width=\linewidth]{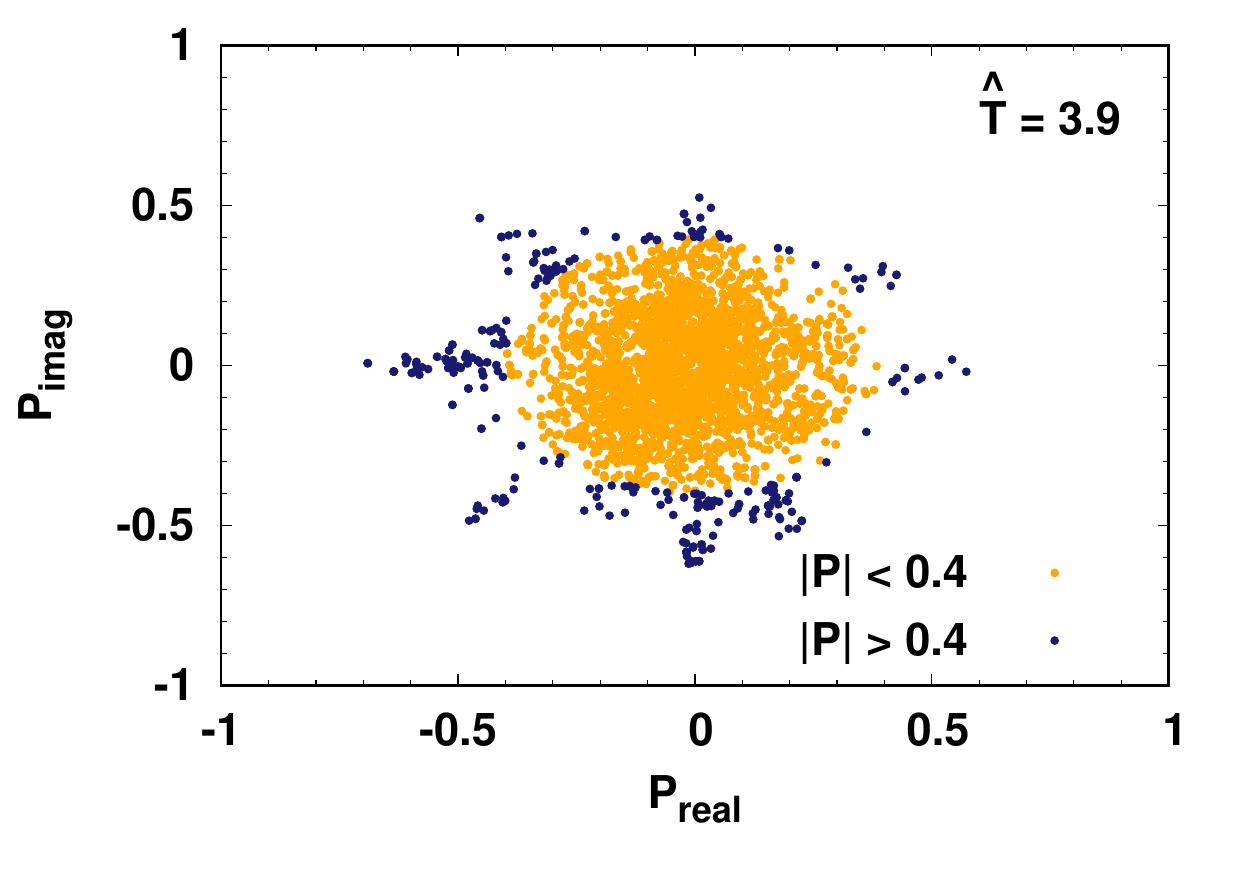}
\caption{$\That = 3.9$}
\end{subfigure}
\begin{subfigure}{0.475\textwidth}
\includegraphics[width=\linewidth]{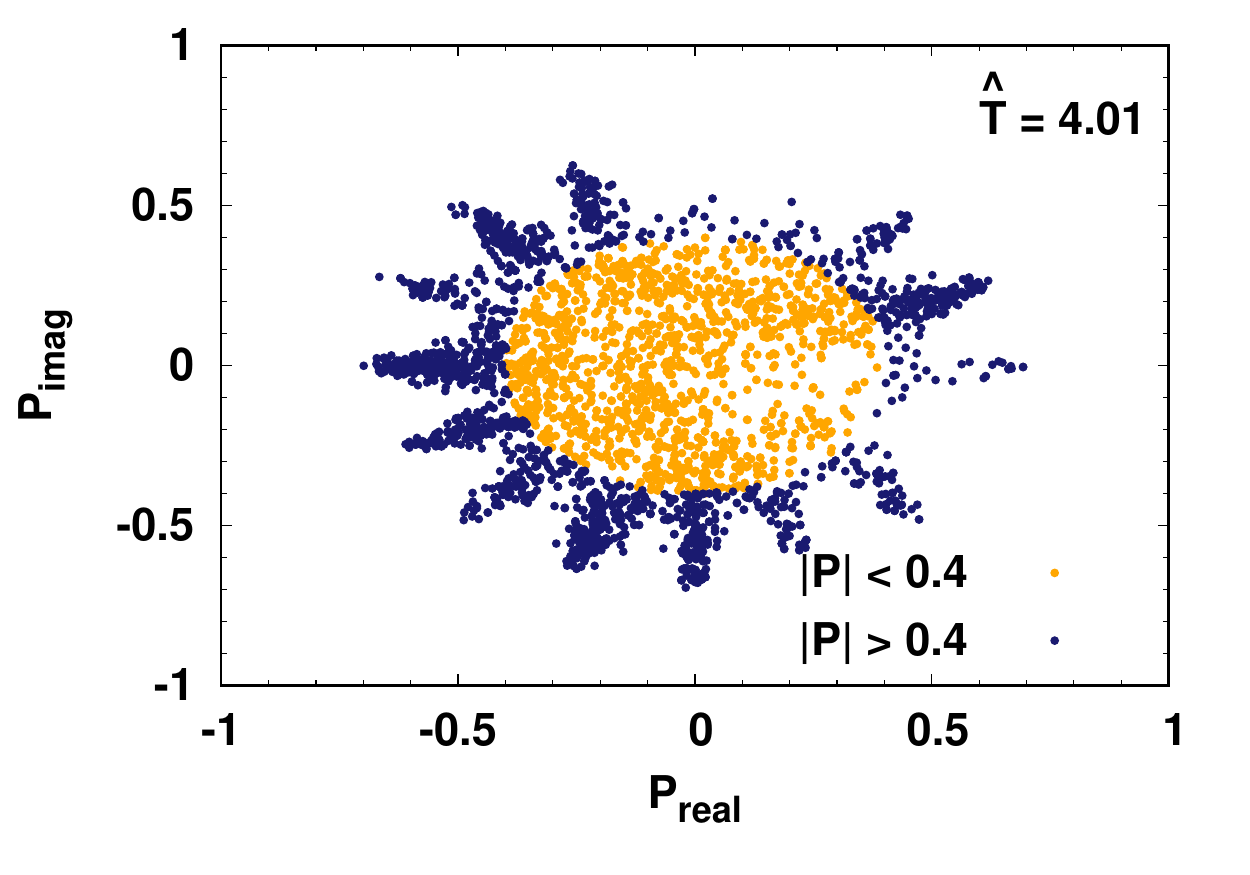}
\caption{$\That = 4.01$}
\end{subfigure}
\begin{subfigure}{0.475\textwidth}
\includegraphics[width=\linewidth]{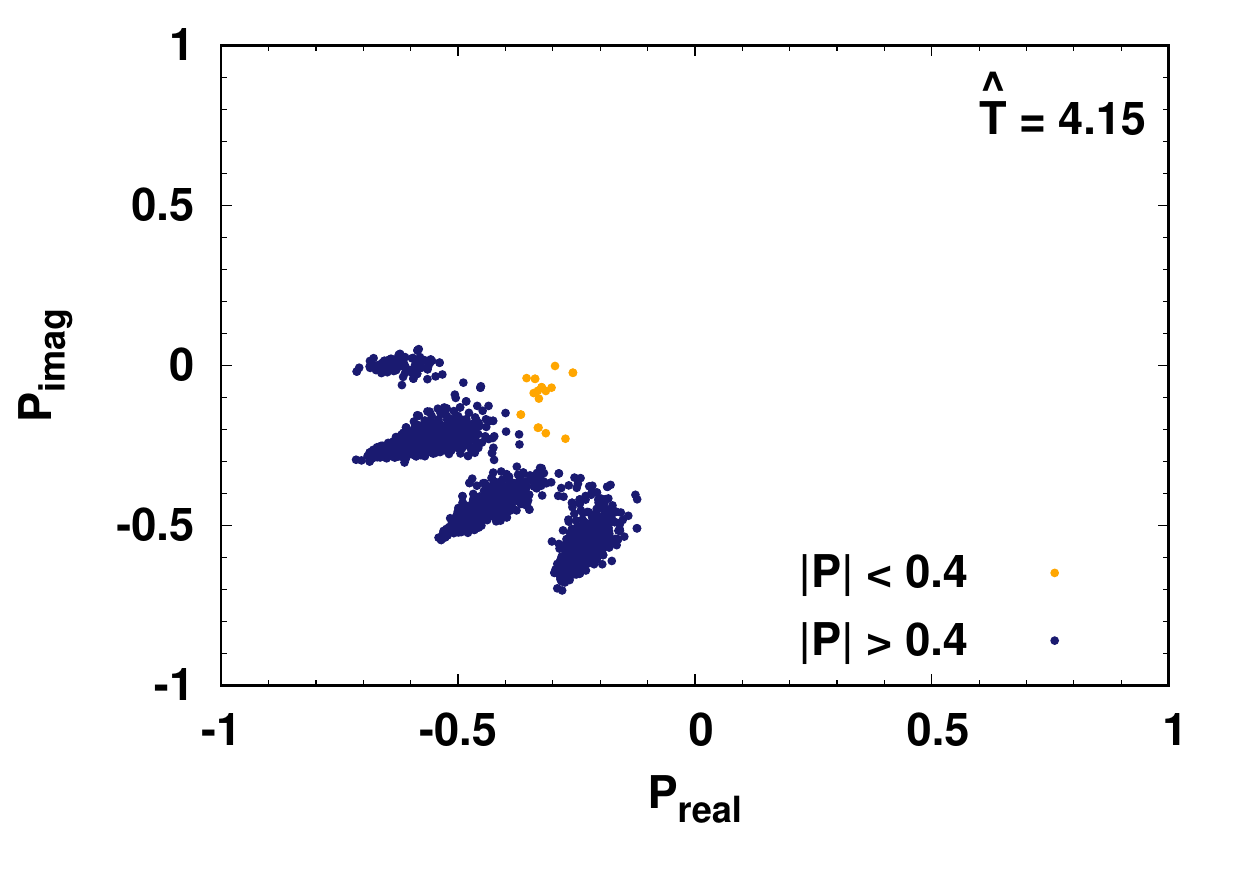}
\caption{$\That = 4.15$}
\end{subfigure}
\begin{subfigure}{0.475\textwidth}
\includegraphics[width=\linewidth]{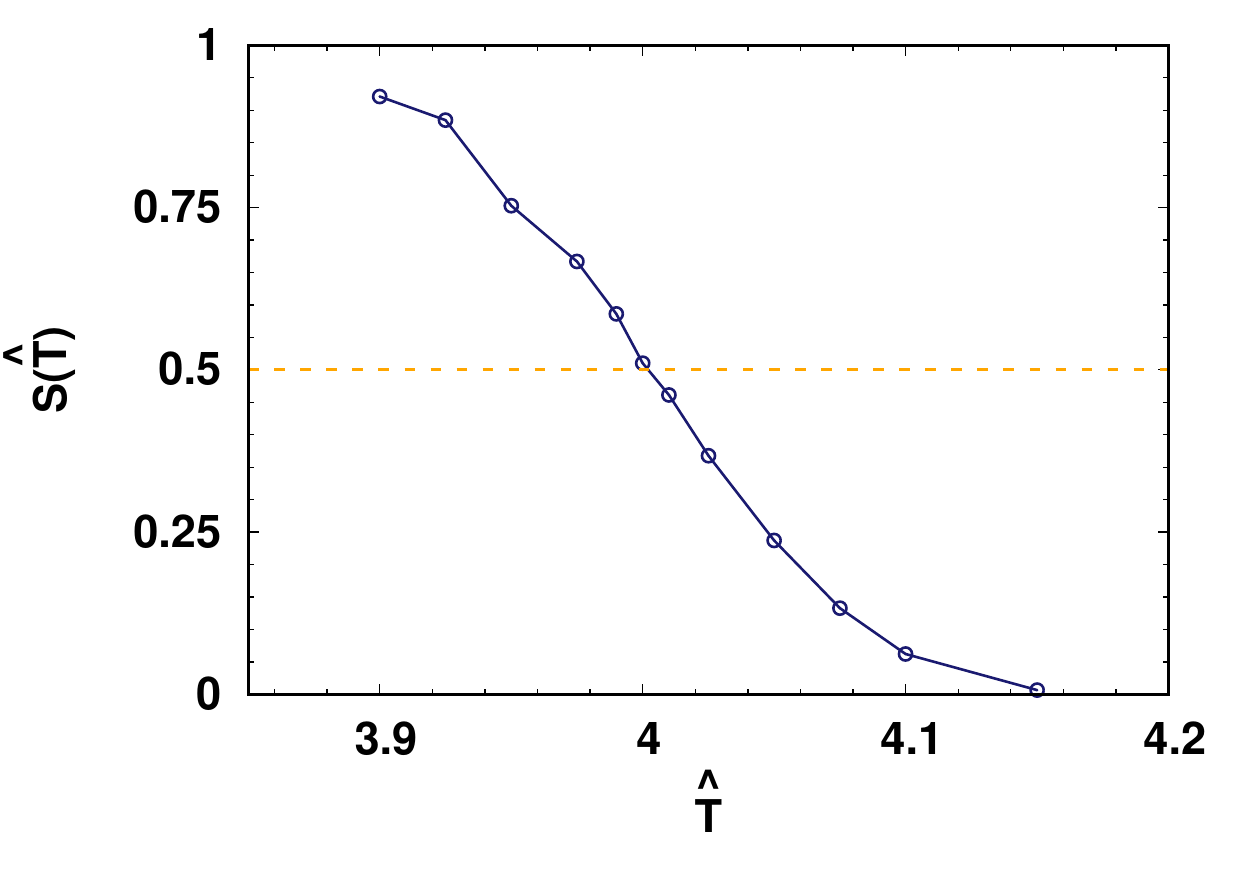}
\caption{$S(\That)$}
\end{subfigure}
\caption{\label{fig:scatter_separatrix_1}Polyakov loop scatter plots for $N = 16$ with $\muhat = 44.66$ for three values of \That and the same $r_S = 0.4$ circular separatrix as \fig{fig:scatter_separatrix_2}.  The resulting $S(\That)$ in panel (d) identifies the critical temperature $\Thatc \approx 4$.}
\end{figure}

We treat the radius $r_S$ as an adjustable parameter,\footnote{In \refcite{Francis:2015lha}, the separatrix is defined using the position of the minimum between two peaks in the distribution of Polyakov loop measurements.} which introduces a systematic uncertainty from our choice of $r_S$.
For all \muhat we consider, we find $r_S = 0.4$ provides stable and reliable results.
We also observe that the systematic dependence on our choice of $r_S$ becomes less significant as \muhat increases.
In Figs.~\ref{fig:scatter_separatrix_2} and \ref{fig:scatter_separatrix_1} we show representative Polyakov loop scatter plots, separatrices, and the resulting $S(\That)$ for $\muhat = 1$ with $N = 32$ and $\muhat = 44.66$ with $N = 16$, respectively.

\begin{figure}[tbp]
\centering
\begin{subfigure}{0.6\textwidth}
\includegraphics[width=\linewidth]{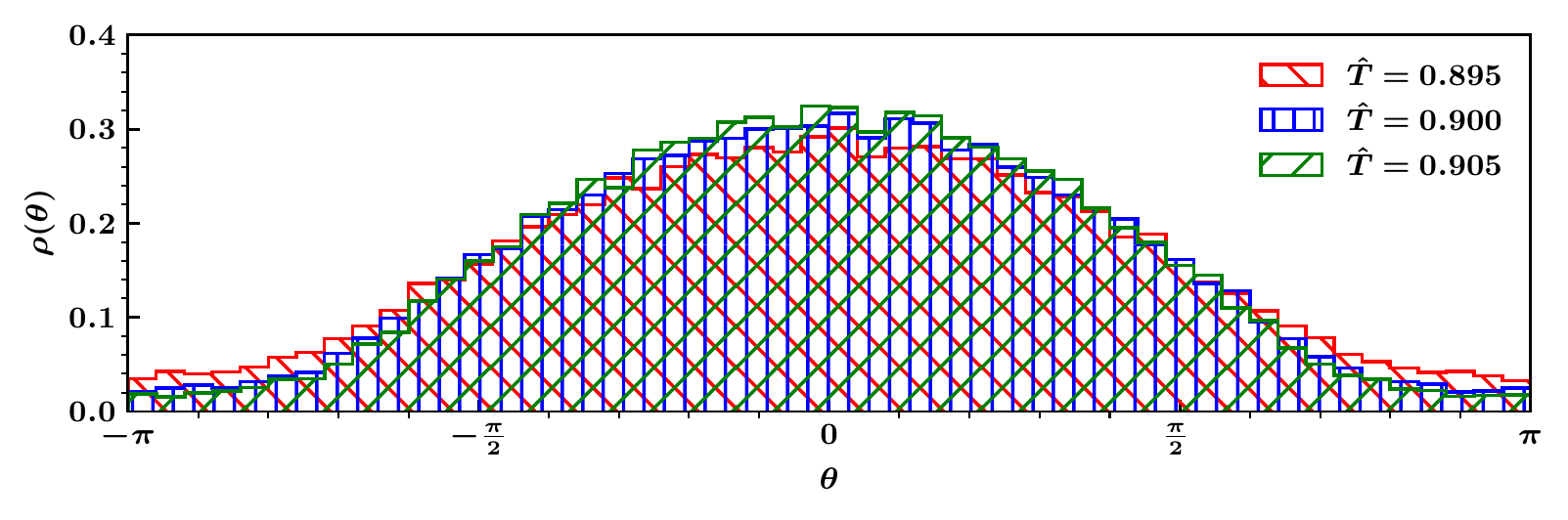}
\caption{$N = 16$}
\end{subfigure}
\begin{subfigure}{0.6\textwidth}
\includegraphics[width=\linewidth]{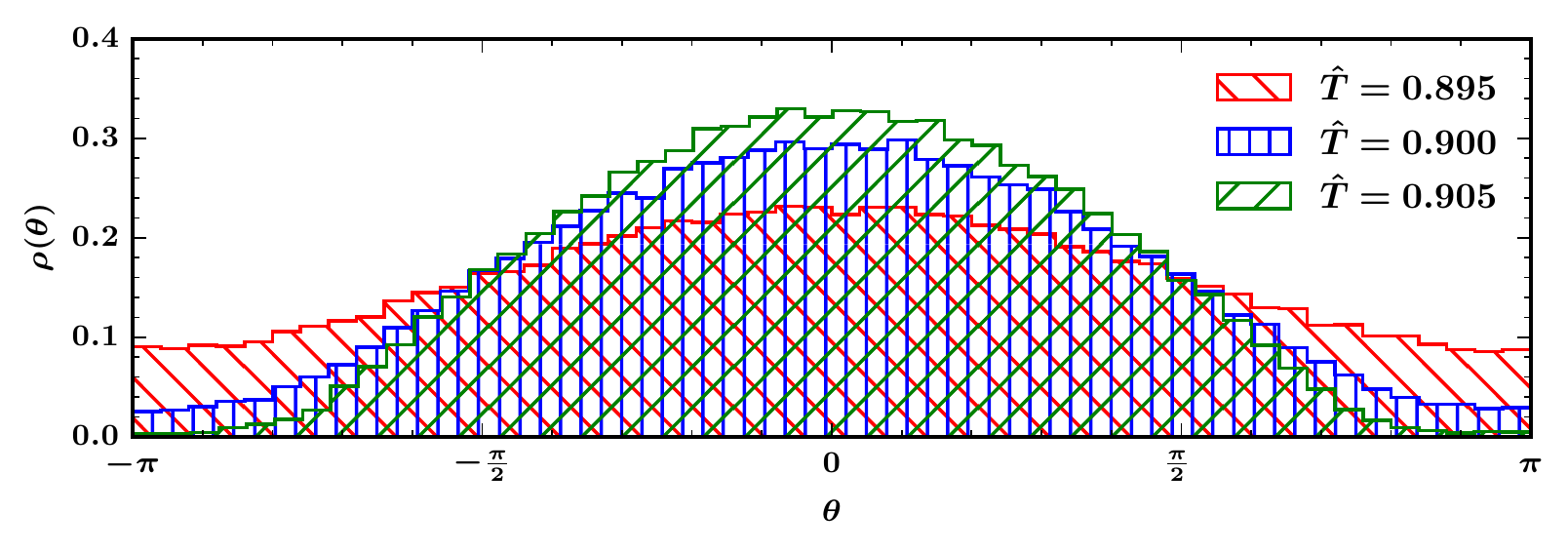}
\caption{$N = 32$}
\end{subfigure}
\begin{subfigure}{0.6\textwidth}
\includegraphics[width=\linewidth]{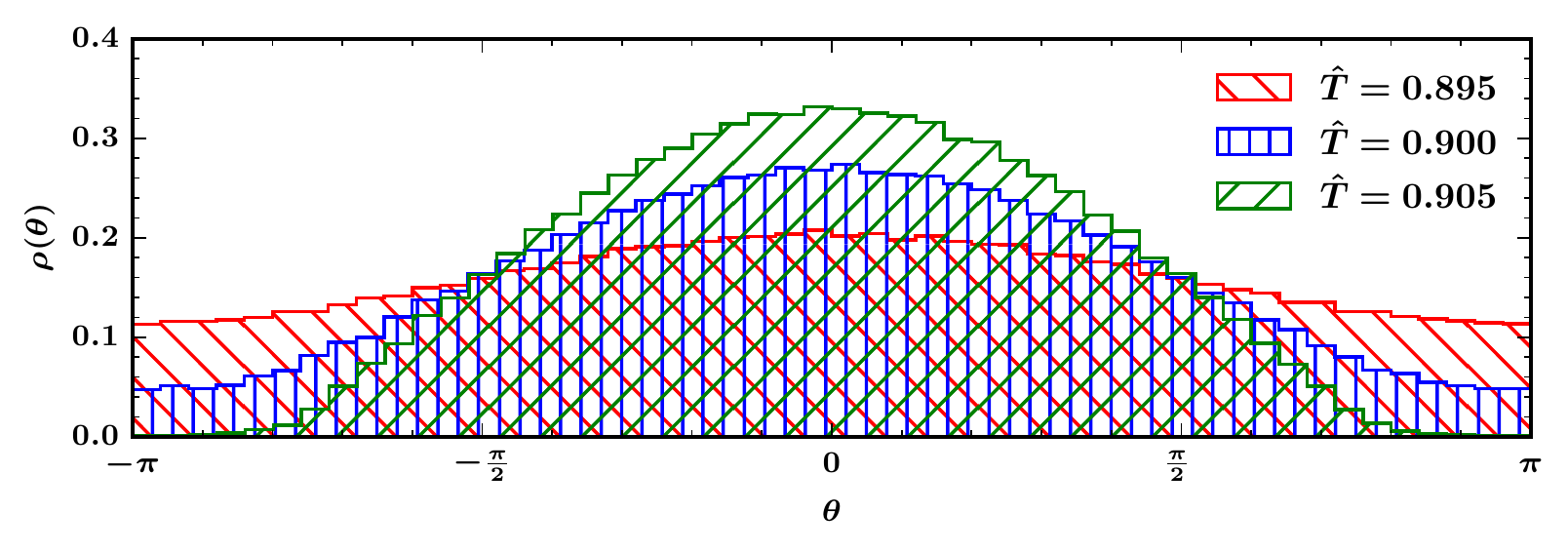}
\caption{$N = 48$}
\end{subfigure}
\caption{\label{fig:Dist}Angular distributions of Polyakov loop eigenvalues for $\muhat = 0.5$ with $N = 16$, $32$ and $48$, considering three temperatures around the critical $\Thatc = 0.900(2)$.  A gap appears for $\That = 0.905 > \Thatc$, while the distribution for $\That = 0.895 < \Thatc$ becomes more uniform as $N$ increases.}
\end{figure}

The $N$ eigenvalues of the Polyakov loop provide yet another means both to estimate the critical temperature and to characterize the phases between which the system transitions.
Deep in the confined phase, the angular distribution of these eigenvalues is uniform around the unit circle, while deep in the deconfined phase their distribution is localized around some angle, which we can set to $\theta = 0$ by convention.
This behavior can be modeled as
\begin{equation}
  \rho(\theta) = \frac{1}{2 \pi} + \frac{1}{q \pi} \cos \theta \qquad\qquad -\pi \leq \theta < \pi,
\end{equation}
where the positive parameter $q \to \infty$ in the uniform limit, while a gap opens for $q < 2$.
In \fig{fig:Dist} we show a representative example of this gap opening at the critical \Thatc identified from the Polyakov loop susceptibility and separatrix, confirming that the corresponding transition is between the uniform confined phase and the gapped deconfined phase.
In this figure we consider our most challenging data set with the smallest $\muhat = 0.5$ and three $N = 16$, $32$ and $48$.
Comparing these three values of $N$ allows us to confirm that the transition becomes sharper as $N$ increases: the distribution for $\That = 0.895 < \Thatc$ becomes more uniform for larger $N$ while the size of the gap for $\That = 0.905 > \Thatc$ also increases.

Finally, we also check that the Myers transition and the confinement transition occur at the same critical temperature, by defining
\begin{equation}
  \label{eq:Delta}
  \De \equiv \Bigg|\Thatc(C_{_V})  - \Thatc(\chi)\Bigg|
\end{equation}
to quantify the difference between the locations of the specific heat and Polyakov loop susceptibility peaks.
Our results for \De in \tab{tab:Tc} vanish within uncertainties for all $\muhat < 40$, consistent with the existence of only a single phase transition.
For $\muhat = 44.66$ we do not observe well-defined peaks and rely on the separatrix method to determine $\Thatc$.
If there were separate phase transitions signalled by these observables, in order to be consistent with these results their critical temperatures would need to be too close to resolve with $N \leq 48$.

\subsection{Critical temperature dependence on deformation parameter}

\begin{figure}[tbp]
\centering
\includegraphics[width=0.9\linewidth]{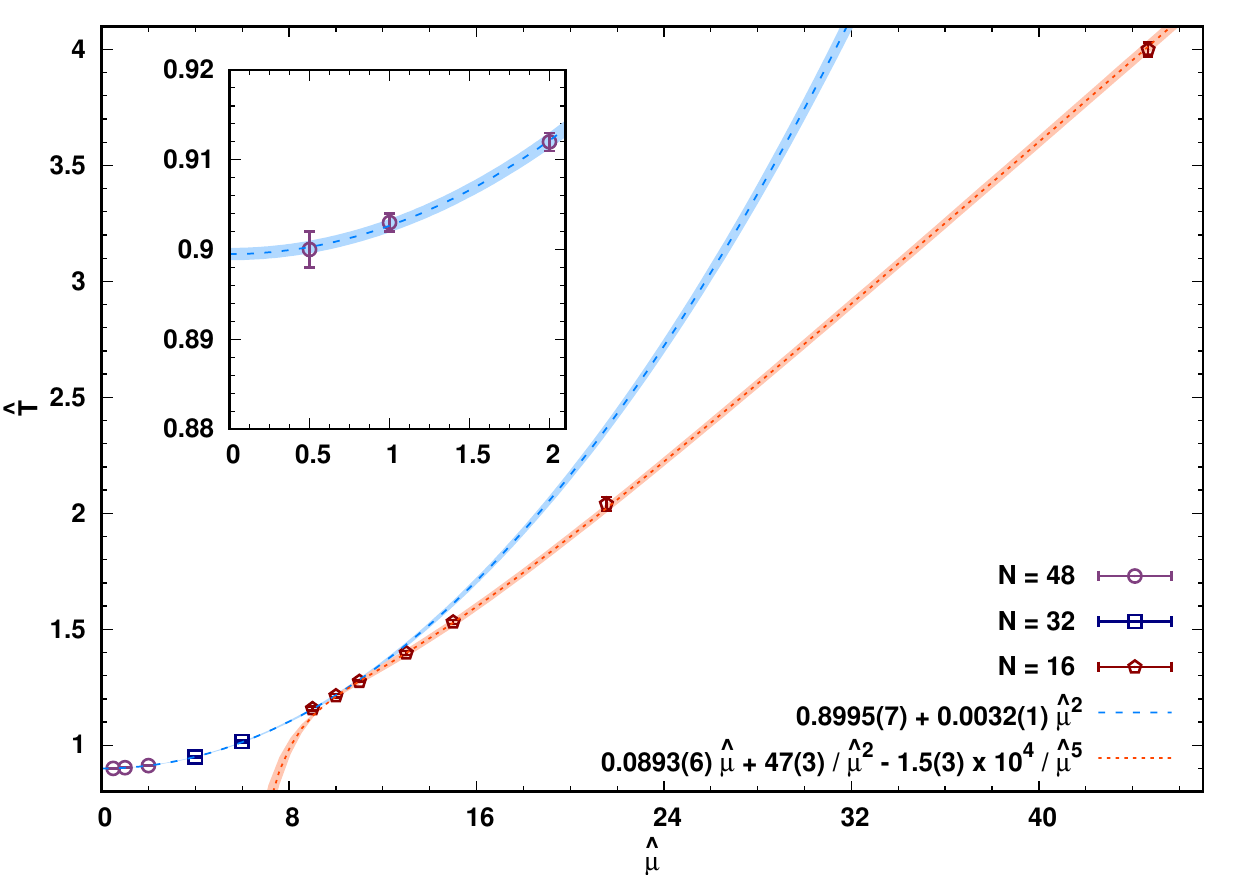}
\caption{\label{fig:tc_mu1}The $\That$--$\muhat$ phase diagram of the bosonic BMN model from our $N_{\tau} = 24$ results for \Thatc computed with $N = 16$, $32$ and $48$.  The red curve shows our fit of the $\muhat \geq 10$ results to \eq{eq:large_fit}, while the blue curve shows our fit of the $\muhat \leq 10$ results to \eq{eq:small_fit}.  Our results self-consistently identify the $\muhat_{\star} \sim 10$ separating the small- and large-$\muhat$ regimes.  The inset zooms in on the $\muhat \to 0$ limit.}
\end{figure}

Figure~\ref{fig:tc_mu1} plots our critical temperature results obtained above, to visualize the phase diagram of the bosonic BMN model in the $\Thatc$--$\muhat$ plane.
We now analyze the dependence of \Thatc on the deformation parameter.
Considerations of the simplified $\muhat \to \infty$ and $\muhat \to 0$ limits suggest that this dependence must differ for large vs.\ small $\muhat$.
Our results confirm this, and also identify the $\muhat_{\star} \sim 10$ separating these two regimes.

First, for large \muhat the expected critical temperature is easy to calculate using the argument~\cite{Aharony:2003sx, Furuuchi:2003sy} that for a model with $D > 1$ matrices (bosonic or fermionic), of masses $\om_j > 0$ with $j = 1, \cdots, D$, the inverse critical temperature $\be_{\text{c}}$ is given by the solution of
\begin{equation}
  \sum_{j = 1}^D e^{- \be \om_j} = 1.
\end{equation}
For the case of the bosonic BMN model with $D = 3 + 6$, the large-$\mu$ critical temperature is the solution of
\begin{equation}
3 e^{ - \be \mu / 3} + 6 e^{ - \be \mu / 6} - 1 = 0,
\label{eqn:T/mu}
\end{equation}
which gives
\begin{equation}
  \label{eq:free}
  \frac{1}{\mu \be_{\text{c}}} = \frac{\Tc}{\mu} = \frac{\Thatc}{\muhat} = \frac{1}{6 \ln ( 3 + 2 \sqrt{3} )} = 0.089305\dots
\end{equation}

A similar analysis can be done for the full BMN model, which reduces to a supersymmetric Gaussian model in the $\muhat \to \infty$ limit.
In this case, Refs.~\cite{Furuuchi:2003sy, Spradlin:2004sx, Hadizadeh:2004bf} have perturbatively computed the critical temperature of the confinement transition up to next-to-next-to-leading order in $1/\muhat^3 \ll 1$:
\begin{equation}
  \label{eq:perturbative_mu}
    \Thatc = \frac{\muhat}{12 \ln 3} \Bigg[1 + \frac{320}{3} \frac{1}{\muhat^3} - \left( \frac{458321}{12} + \frac{1765769 \ \ln 3}{144} \right) \frac{1}{\muhat^6} + \cO\left(\frac{1}{\muhat^9}\right) \Bigg].
\end{equation}
In the $\muhat \to \infty$ limit, this produces a smaller $\Tc/\mu \approx 0.076$ compared to the bosonic BMN result in \eq{eq:free}.
Motivated by the functional form of this perturbative result, we adopt the following ansatz to fit our \Thatc results for sufficiently large $\muhat$:
\begin{equation}
  \label{eq:large_fit}
  \Thatc = \muhat \left[C + H \frac{1}{\muhat^3} + F \frac{1}{\muhat^6}\right],
\end{equation}
where we expect $C = 1/[6 \ln ( 3 + 2 \sqrt{3} )] \approx 0.0893$ from \eq{eq:free}.
The fit to our results for $\muhat \geq 10$ shown in \fig{fig:tc_mu1} indeed produces $C = 0.0893(6)$, providing a good check of our numerical setup and code.

Also from \fig{fig:tc_mu1} we can observe that this ansatz fails for $\muhat_{\star} \lesssim 10$, consistent with the expected breakdown of the perturbative expansion when the coupling $1 / \muhat^3$ is too large.
In the small-$\muhat$ regime we must use a different fit form to describe the dependence of the critical temperature on the deformation parameter.
While it might be possible to carry out a strong-coupling expansion in this regime, for the purposes of this work we will simply employ an empirical expansion in powers of $\muhat^2$.
The recent \refcite{Bergner:2021goh} takes the same approach, employing the quadratic ansatz
\begin{equation}
  \label{eq:small_fit}
  \Thatc = A + B \muhat^2,
\end{equation}
where $A$ is the constant critical temperature of the bosonic BFSS model.

We will also use \eq{eq:small_fit} to fit our \Thatc results for small $\muhat$.
Using our conventions, \refcite{Bergner:2021goh} reports $A = 0.8846(1)$ and $B = 0.00330(2)$ from a fit to their critical temperature results for $0.375 \leq \muhat \leq 3$ (i.e., $0.125 \leq \mu \leq 1$ in their conventions).
The fit to our results for $0.5 \leq \muhat \leq 10$ shown in \fig{fig:tc_mu1} produces $A = 0.8995(7)$ and $B = 0.0032(1)$, with purely statistical uncertainties.
While our result for $B$ agrees with \refcite{Bergner:2021goh}, there is a clear tension in $A$.
The inset in \fig{fig:tc_mu1} makes it clear that our numerical results demand $A > 0.89$ regardless of the range of \muhat we include in our fit.
Finite-$N$ and discretization artifacts could play a role in this disagreement.
So far we have considered only $N \leq 48$ rather than the $N \leq 64$ that \refcite{Bergner:2021goh} was able to reach.
Although we use the same $N_{\tau} = 24$ as \refcite{Bergner:2021goh}, our first-order lattice finite-difference operator in \eq{eq:lat_act} differs from the second-order discretization they employ.
The lattice action of \refcite{Bergner:2021goh} also includes a Faddeev--Popov term from gauge fixing to the static diagonal gauge, though we do not expect this gauge fixing to affect the critical temperature.

\begin{table}[btp]
\centering
\vspace{0.5cm}
\begin{tabular}{ c | l | l }
\hline
\hline
Coefficient & Value & Reference \\
\hline
$A$ & $0.8846(1)$ & \refcite{Bergner:2021goh} \\
$A$ & $0.8995(7)$ & This work \\ \hline
$B$ & $0.00330(2)$ & \refcite{Bergner:2021goh} \\
$B$ & $0.0032(1)$ & This work \\ \hline
$C$ & $0.0893$ & Refs.~\cite{Aharony:2003sx, Furuuchi:2003sy} \\
$C$ & $0.0893(6)$ & This work \\ \hline
$H$ & $47(3)$ & This work \\ \hline
$F$ & $-15(3) \times 10^3$ & This work \\
\hline
\end{tabular}
\caption{\label{tab:coeffs}Comparison of the values of the fit parameters appearing in Eqs.~\ref{eq:large_fit} and \ref{eq:small_fit}.}
\end{table}

\tab{tab:coeffs} summarizes our findings for the coefficients in Eqs.~\ref{eq:large_fit} and \ref{eq:small_fit}.
We note that $H$ and $F$ are new predictions from this work.
Comparing these with the perturbative computation for the full BMN model in \eq{eq:perturbative_mu}, we see that our non-perturbative results for the bosonic case are both a few times larger: $H_{\text{BBMN}} \simeq 47$ compared to $H_{\text{BMN}} \simeq 8$, while $F_{\text{BBMN}} \simeq - 15\times 10^3$ compared to $F_{\text{BMN}} \simeq -4\times 10^3$.

\subsection{Order of the phase transition}

\begin{figure}[tbp]
\centering
  \begin{subfigure}{0.49\textwidth}
    \includegraphics[width=\linewidth]{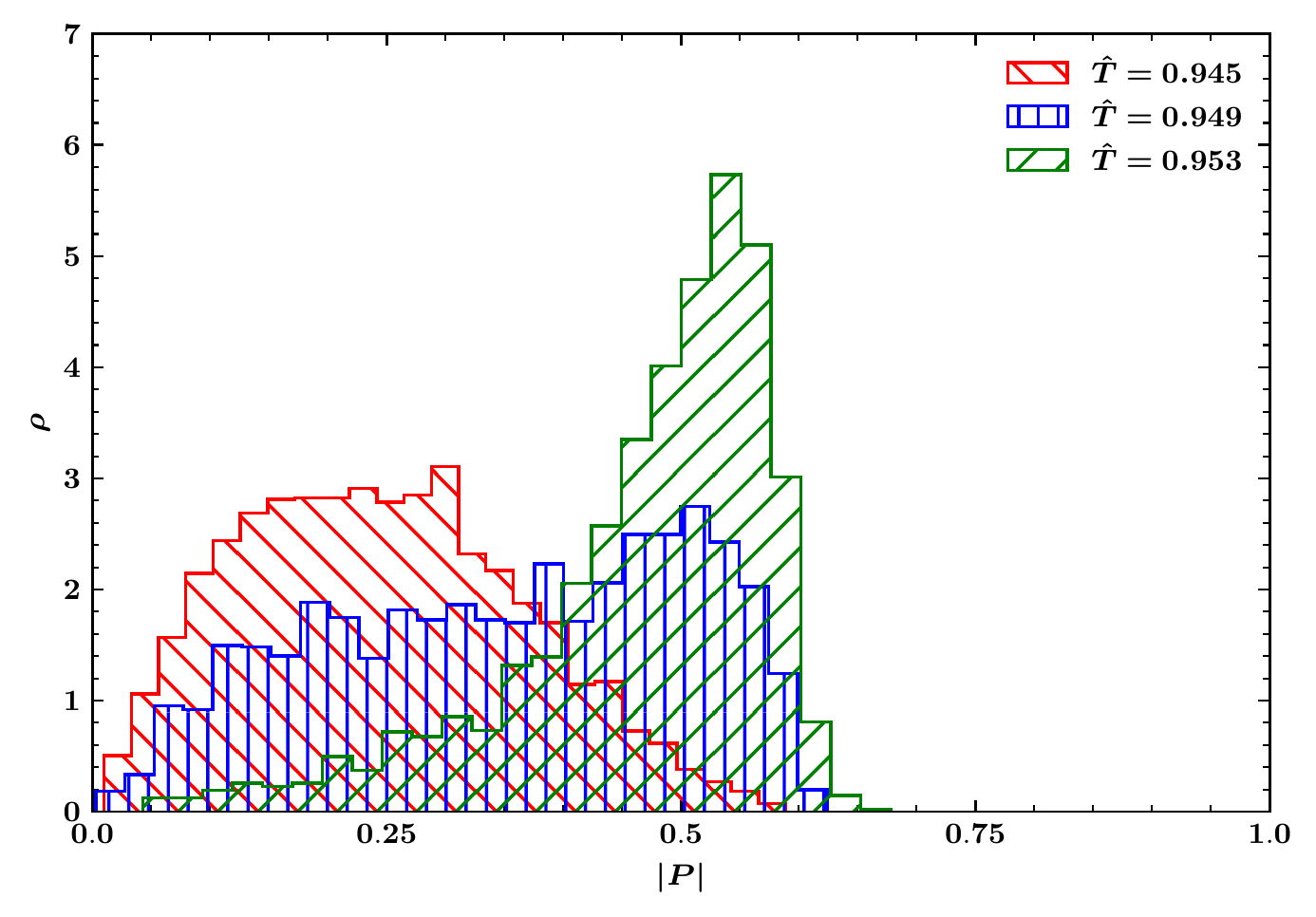}
    \caption{Polyakov loop magnitude distributions}
  \end{subfigure}
  \begin{subfigure}{0.49\textwidth}
    \includegraphics[width=\linewidth]{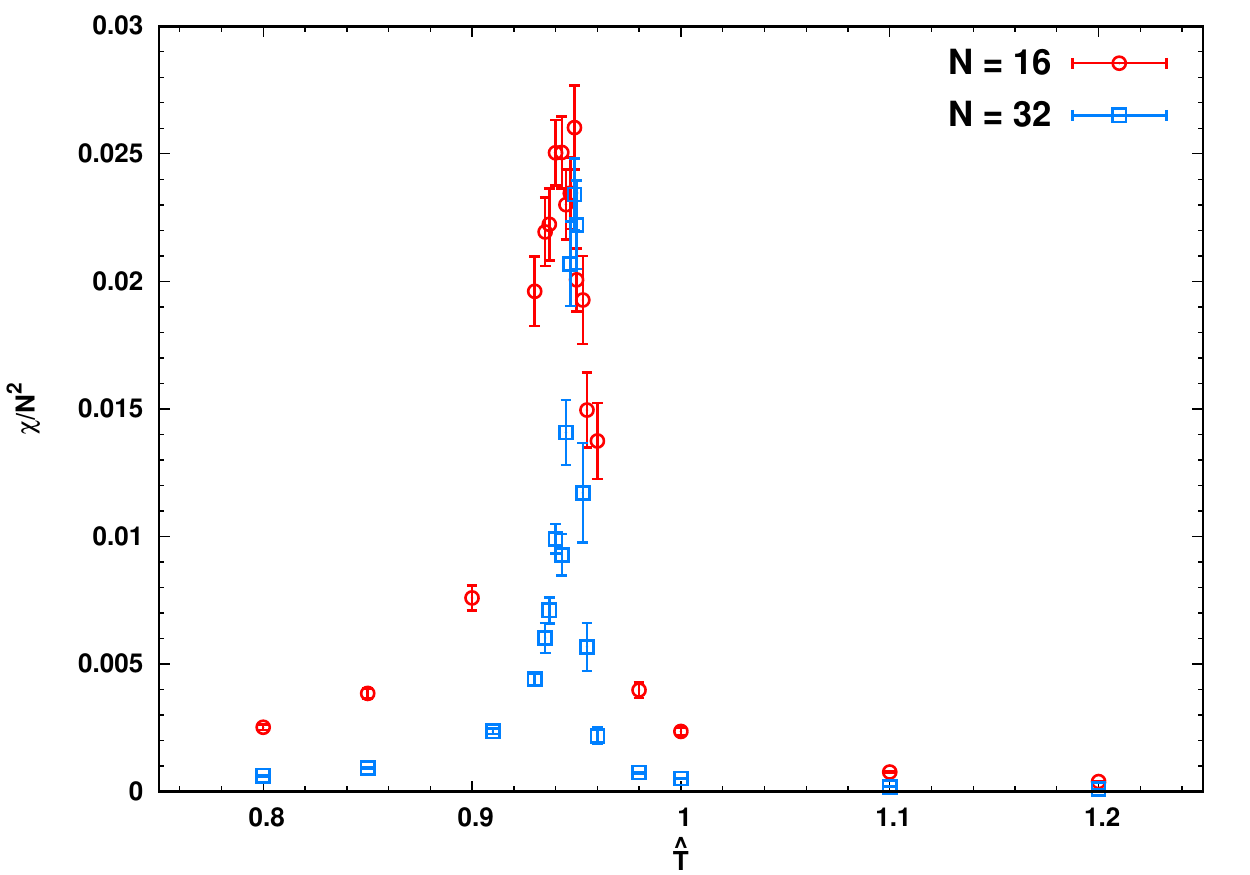}
    \caption{Polyakov loop susceptibilities $\chi/N^2$ vs.\ \That}
  \end{subfigure}
  \caption{\label{fig:transition_order}Signs that the phase transition is first order, for $\muhat = 4$.  (a) The distribution of the Polyakov loop magnitude has support across two different regions for $\Thatc \approx 0.949$, suggesting a developing two-peak structure. (b) Normalizing the Polyakov loop susceptibility by $N^2$ produces the same peak height for $N = 16$ and $32$, indicating that the maximum $\chi$ is scaling with the number of degrees of freedom $\propto N^2$.}
\end{figure}

We expect that the single phase transition we observe is first order.
We confirm this in two ways.
First, in the left panel of \fig{fig:transition_order} we plot the distribution of the Polyakov loop magnitude for three temperatures around the critical $\Thatc \approx 0.949$, for a representative $\muhat = 4$ with $N = 32$.
While the $|P|$ distributions for $\That = 0.945$ and $\That = 0.953$ each have a single peak in two different regions, respectively corresponding to the confined and deconfined phases, the distribution for $\That = 0.949$ has support across both of these regions.
This suggests that a developing two-peak structure would be visible for larger $N > 32$, which is characteristic of phase coexistence at a first-order transition.

Second, in the right panel of \fig{fig:transition_order} we check the scaling of the Polyakov loop susceptibility with $N$.
Because the thermodynamic limit for the bosonic BMN matrix model corresponds to $N^2 \to \infty$, this scaling can distinguish between first- and higher-order phase transitions~\cite{Fukugita:1990vu, Azuma:2014cfa}.
Considering the same $\muhat = 4$, we plot $\chi / N^2$ against $T$ for both $N = 16$ and $32$.
Within uncertainties, both values of $N$ produce the same peak height, again suggesting a first-order transition where the maximum $\chi$ would scale with the number of degrees of freedom.

\subsection{Dependence of the internal energy on \That and $\muhat$}\label{sec:energy}

Finally, we comment on the internal energy of the bosonic BMN model.
Let us begin in the $\muhat \to \infty$ limit where this system reduces to a gauged Gaussian model with $D = 9$ scalar matrices.
For general $D$, \refcite{Kabat:1999hp} computed that the internal energy of this Gaussian model is
\begin{equation}
  \label{eq:energy_T}
  \frac{1}{N^2} \widehat{E} = \frac{3}{4} (D - 1) \That + \cO\Big(\frac{1}{N^2}\Big).
\end{equation}
for high temperatures and large $N$.
Plugging in the $D = 9$ relevant for the bosonic BMN model, we want to explore how this result will be modified for finite $\muhat$.
We expect that the relevant parameter will be $\frac{\That}{\muhat} = \frac{T}{\mu}$, so that
\begin{equation}
  \frac{1}{N^2} \widehat{E} = 6\That \Big[1 + f \left(\frac{T}{\mu} \right) \Big] + \cO\Big(\frac{1}{N^2}\Big)
\end{equation}
for some as-yet unknown function $f$. 

\begin{figure}[tbp]
\centering
  \includegraphics[width=0.9\linewidth]{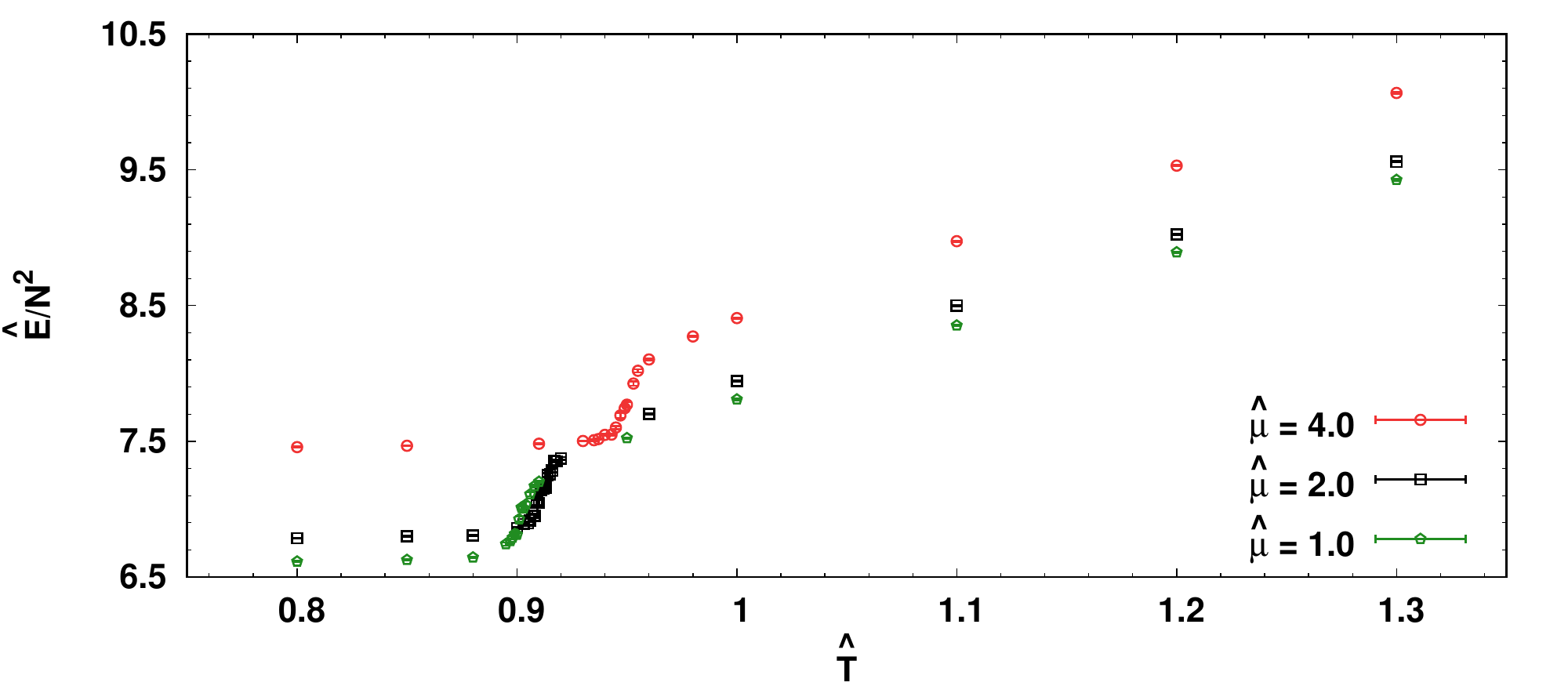}
  \caption{\label{fig:Energy_ramp}Internal energy vs.\ temperature for various \muhat values and $N = 32$. We see that for $\That < \Thatc$, the energy has no temperature dependence, and after transition it depends linearly on temperature.}
\end{figure}

In \fig{fig:Energy_ramp} we show our lattice results for the temperature dependence of the energy, for $\muhat = 1$, $2$ and $4$ with $N = 32$.
Although our lattice calculations focus on the transition regions, we do consider enough $\That > \Thatc$ points in the deconfined phase to clearly see the leading-order linear dependence predicted by \eq{eq:energy_T}.
This is in contrast to the $\That$-independent energy in the $\That < \Thatc$ confined phase.
Although the energy depends on \muhat in both phases, we observe that the high-temperature slope is insensitive to the deformation parameter within the range $1 \leq \muhat \leq 4$.
A precise determination of $f(T/\mu)$ will be interesting to pursue through future generations of bosonic BMN lattice calculations.

\section{Conclusion}
\label{sec:summary}

We have presented a lattice study of the non-perturbative phase structure of the bosonic BMN matrix model.
Our main results for the transition temperatures \Thatc for twelve $0.5 \leq \muhat \leq 44.66$, collected in \fig{fig:tc_mu1}, show how our numerical investigations smoothly connect the bosonic BFSS model in the $\muhat \to 0$ limit to the known behavior of the $\muhat \to \infty$ gauged Gaussian model.
In addition to monitoring several standard observables and susceptibilities, we have also applied a novel separatrix method to determine these critical temperatures.
We observed only a single transition, finding evidence that it is first order, and analyzing the Polyakov loop eigenvalues to confirm that it separates the uniform confined phase from the gapped deconfined phase.

Using our results for $\Thatc$, we have investigated the functional forms of the dependence of the critical temperature on the deformation parameter in both the small- and large-$\muhat$ regimes.
Our results for the parameters of these functional forms, collected in \tab{tab:coeffs}, agree with some existing values in the literature, and also provide some new predictions.
However, there is a disagreement with results from Refs.~\cite{Bergner:2019rca, Bergner:2021goh} for \Thatc in the $\muhat \to 0$ bosonic BFSS limit, which deserves further investigation.

For these future investigations, we are particularly interested in exploring smaller $\muhat$, which will require larger $N > 48$ to overcome challenges associated with flat directions and metastable vacua that make the numerical calculations more difficult.
We also have lattice investigations of the full BMN model underway~\cite{Schaich:2022duk}, with similar plans to pursue smaller \muhat with larger $N$.

We have already mentioned our ambition to determine the function $f(T/\mu)$ that modifies the dependence of the internal energy on the temperature and deformation parameter in \secref{sec:energy}.
In that section we also raised the possibility of generalizing the bosonic BMN model to a different number of scalar matrices, $D \neq 9$.
It would be interesting to explore how the phase diagram and the order of phase transitions depend on $D$.
\refcite{Morita:2020liy} recently addressed this problem for the analogous generalization of the $\muhat = 0$ bosonic BFSS model, investigating that system for a range of $D$ and concluding that the transition changes from first order to second order for $D \sim 36$.
In the future we hope to report how the BMN deformation affects this phenomenon.

Finally, we also plan to study the `ungauged' version of the BMN matrix model, with and without fermions, again building on prior investigations of the $\muhat = 0$ BFSS model~\cite{Maldacena:2018vsr, Berkowitz:2018qhn}.
In addition to exploring the effects of the deformation parameter in this context, it will be interesting to see the extent to which holographic arguments carry over to the non-supersymmetric bosonic BMN model.

\acknowledgments

We thank Denjoe O'Connor and Samuel Kov\'{a}\v{c}ik for their comments during a seminar by RGJ at the Dublin Institute for Advanced Studies (DIAS) in June 2021. NSD thanks the Council of Scientific and Industrial Research (CSIR), Government of India, for the financial support through a research fellowship (Award No.~{09/947(0119)/2019-EMR-I}). The work of AS was partially supported by an INSPIRE Scholarship for Higher Education by the Department of Science and Technology, Government of India. RGJ is supported by postdoctoral fellowship at the Perimeter Institute for Theoretical Physics. Research at Perimeter Institute is supported in part by the Government of Canada through the Department of Innovation, Science and Economic Development Canada and by the Province of Ontario through the Ministry of Colleges and Universities. The work of AJ was supported in part by the Start-up Research Grant (No.~{SRG/2019/002035}) from the Science and Engineering Research Board (SERB), Government of India, and in part by a Seed Grant from the Indian Institute of Science Education and Research (IISER) Mohali. DS was supported by UK Research and Innovation Future Leader Fellowship {MR/S015418/1} and STFC grant {ST/T000988/1}. Numerical calculations were carried out at the University of Liverpool and IISER Mohali.

\appendix

\section{Internal energy on the lattice}
\label{sec:int_energy}

To obtain the expression for the internal energy, we consider deforming the partition function by a small amount from $Z$ to $Z'$, which is expressed by the following set of transformations:
\begin{align}
  \label{eqn:transformat}
  t' & = \frac{\be'}{\be} t, &
  A'\left(t'\right) & = \frac{\be}{\be'} A(t), &
  X_i'\left(t'\right) & = \sqrt{\frac{\be'}{\be}} X_i(t).
\end{align}
Note that we have $\left[D X'\right] = [D X]$ and $\left[D A'\right] = [D A]$.
For the bosonic BMN model, it is convenient to break up the action \eq{eq:action} into two pieces:
\begin{align}
  S & = S_0 + S_{\mu}, \label{eq:act_orig} \\
  S_0 & = \frac{N}{4 \la} \int_0^\be d\tau \ \mbox{Tr} \Bigg( -(D_{\tau} X_i)^2 - \frac{1}{2} \sum_{i < j} [X_i, X_j]^2 \Bigg), \\
  S_{\mu} & = - \frac{N}{4 \la} \int_0^{\be} d\tau \ \mbox{Tr} \Bigg( \frac{\mu^2}{9} X_I^2 + \frac{\mu^2}{36} X_A^2 - \frac{\sqrt{2} \mu}{3} \eps_{IJK} X_I X_J X_K  \Bigg).
\end{align}
Applying Eqs.~\eqref{eqn:transformat}, and defining $\De\be \equiv \be' - \be$, we find
\begin{align}
  S' & = S_0' + S_{\mu}', \\
  S_0' & = S_0 + \frac{N}{4 \la} \int_0^\be d\tau \ \mbox{Tr}\Bigg( - \frac{3}{2} \frac{\De \be}{\be} \sum_{i < j} [X_i, X_j]^2 \Bigg) + \cO(\De \be^2), \\
  S_{\mu}' & = S_\mu - \frac{N}{4 \la} \int_0^\be d\tau \ \mbox{Tr}\Bigg( 2 \frac{\De \be}{\be} \left(\frac{\mu}{3} X_I \right)^2 + 2 \frac{\De \be}{\be} \left( \frac{\mu}{6} X_A\right)^2 \nn \\
  & \hspace{6.5 cm} - \frac{5 \sqrt{2} \mu}{6} \frac{\De \be}{\be} \eps_{IJK} X_I X_J X_K \Bigg) + \cO(\De \be^2). \label{eq:act_mod}
\end{align}
The partition function therefore becomes
\begin{align}
  Z(\be') & = \int [D X']_{\be'} [D A']_{\be'} \ e^{-S'} = \int [D X]_{\be} [D A]_{\be} \ e^{- S} e^{E \De \be + \cO(\De\be^2)} \nn \\
  & = Z(\be) \left[1 + E \De \be + \cO(\De \be^2) \right].
\end{align}
Hence, we can write
\begin{equation}
  \label{eqn:E/N}
  \frac{\widehat{E}}{N^2} = \frac{E}{\la^{1/3} N^2} = \frac{1}{Z(\be) \la^{1/3} N^2} \lim_{\De \be \to 0} \frac{Z(\be') - Z(\be)}{\De \be}.
\end{equation}
Using Eqs.~\eqref{eq:act_orig}--\eqref{eq:act_mod} in \eq{eqn:E/N} we get
\begin{equation}
  \begin{split}
    \frac{\widehat{E}}{N^2} & = \frac{1}{\la^{1/3} \be} \Bigg \langle \frac{1}{4N \la} \int_0^\beta d \tau \ \mbox{Tr} \Bigg( - \frac{3}{2} \sum_{i < j} [X_i, X_j]^2 - 2 \left( \frac{ \mu }{3} X_I \right)^2 - 2 \left( \frac{\mu}{6} X_A \right)^2 \\
    & \hspace{7.5 cm} + \frac{5 \sqrt{2} \mu}{6} \epsilon_{IJK} X_I X_J X_K \Bigg) \Bigg\rangle.
  \end{split}
\end{equation}

Upon discretizing the bosonic BMN model, as discussed in \secref{sec:lat_for}, the dimensionless lattice internal energy takes the form reported in \eq{eqn:E_b}:
\begin{equation}
  \begin{split}
    \frac{\widehat{E}}{N^2} & = \frac{1}{4 N \lalat^{4/3} N_\tau} \Bigg \langle \sum_{n = 0}^{N_\tau - 1} \mbox{Tr} \Bigg( - \frac{3}{2} \sum_{i < j} [X_i, X_j]^2 - \frac{2\mu_{\rm lat}^2}{9} X_I^2 - \frac{\mu_{\rm lat}^2}{18} X_A^2 \\
    & \hspace{6.5 cm} + \frac{5 \sqrt{2} \mu_{\rm lat}}{6} \epsilon_{IJK} X^I X^J X^K \Bigg) \Bigg \rangle.
  \end{split}
\end{equation}

\raggedright
\bibliographystyle{utphys}
\bibliography{v1}
\end{document}